\documentclass[preprint]{aastex}

\shorttitle{Molecular Clouds and H~II Regions I: S175 }
\shortauthors{Azimlu M., Fich M., M$^{\textrm c}$Coey C.}

\begin{document}

\title{Studies of Molecular Clouds associated with H~II Regions: S175}

\author{Mohaddesseh Azimlu, Michel Fich and Carolyn M$^{\textrm c}$Coey}
\affil{Department of Physics \& Astronomy, University of Waterloo,
    Waterloo, ON, Canada, N2L 3G1}

\begin{abstract}

We are  studying the impact of H~II regions on star formation in their associated molecular clouds.   In this paper we present JCMT R$\times$A molecular line observations of S175 and environs. This is the first within a sample of ten H~II regions and their surrounding molecular clouds selected for our study.  We first make $7'\times 7'$  maps in  $^{12}$CO(2-1), which are used to investigate the structure  of the cloud and to identify individual  clumps. Single point observations were made  in $^{13}$CO(2-1) and CS(5-4) at the peak of the $^{12}$CO(2-1) emission within each clump in order to measure the physical properties of the gas. Densities, temperatures, clump masses, peak velocities, and line widths were measured and calculated using these observations.   We have identified  two condensations (S175A and S175B) in the molecular cloud associated with this H~II region.  S175A is adjacent to the ionization front and is expected to be affected by the H~II region while S175B is too distant to be disturbed.  We compare the structure and gas properties of these two regions to investigate how the molecular gas has been affected by the H~II region.
 S175A  has been heated by the H~II region and partially compressed by the ionized gas front, but contrary to our expectation it  is a quiescent region while S175B is very turbulent and dynamically active. Our investigation for the source of  turbulence in S175B  resulted in the detection of an outflow within this region.

\end{abstract}


\section{Introduction}

Molecular clouds are the birth place of stars, and it is essential to study  the formation, evolution and  fragmentation of molecular clouds in order to understand star formation.  The star formation process in molecular clouds is driven by a variety of  mechanisms and is strongly affected by the presence of high-mass stars. When  massive stars form it is expected that they dominate the formation process of other stars formed later in the same cloud. Heating of the molecular cloud by high-mass stars is likely to inhibit further star formation \citep{Scov87}. On the other hand, compression of the cloud, by the action of stellar winds or from the expansion of an H~II region, will enhance star formation \citep{LW79} or even trigger star formation (via sequential star formation or the collect and collapse process, e.g., Elmegreen \& Lada 1977, Zavagno et al. 2006, Deharveng et al. 2008).  Alternatively, the expansion of an H~II region may even blow the cloud apart  (Elmegreen \& Lada  1976). These different processes not only affect the rate of star 
formation within the cloud, but might also be expected to have some effect on the Initial Mass Function (IMF) of the newly forming stars. For example, the Jeans mass increases with cloud  temperature: consequently a warmer cloud is more stable against collapse and the process of fragmentation  may result  in more massive stars with different star formation rates or efficiencies \citep{BB2005}.

It is instructive, therefore, to study the molecular gas associated with H~II regions. Many studies have investigated  the star formation process adjacent to H~II regions (e.g  Deharveng et al. 2005, Kirsanova et al. 2008, Kerton 2008).
Statistically, it has been shown that the most  luminous proto-stars form in molecular clouds associated with H~II regions \citep{dobashi2001}. Other studies suggest that a large fraction of stars originate in clusters (Li et al. 1997), mostly at the peripheries of H~II regions (e.g. Zavagno et al. 2006, Deharveng et al. 2008).

Embedded clusters are associated with clumps and dense cores within clouds, where there is sufficient gas and dust available to form stars; therefore, the spatial distribution of clumps and cores should reflect the stellar distribution of recent or future star-forming regions.   There is not a well-accepted unique definition of cores and clumps. In this work, we call the whole molecular gas associated with the H~II region {\it the cloud} in which we have resolved  three main distinct regions: S175A, S175B and S175C. We define a {\it clump}  to be  condensed material within each region that forms structures larger than the telescope beam size. Each clump is expected to contain sub-structure and exhibits one or more peaks in $^{12}$CO(2-1) emission: we refer to these simply as {\it peaks} as we can not resolve proto-stellar cores within our maps. The $^{12}$CO(2-1) antenna temperature of the brightest peak within each clump is reported as  the clump's temperature  and is used to calculate other parameters such as column density and mass.

This is the first paper resulting from a study of  ten molecular clouds associated with H~II regions which are selected from the Sharpless catalog of H~II regions (Sharpless 1959).   We have selected objects in the outer Galaxy, primarily along the Perseus arm, in order to minimize confusion with background stars and to provide the best estimate of the kinematic distance (for those objects which have no direct distance determination). S175 is our closest source and therefore provides us with the best linear scale resolution in our sample.   We  used the James Clerk Maxwell Telescope (JCMT) to study how the formation of a massive star in this molecular cloud may impact the formation of future stars.  Clumps within the molecular cloud are identified and their physical characteristics, such as mass, column density and temperature, are measured from properties of $^{12}$CO(2-1) and $^{13}$CO(2-1) emission.   We look for  the influence of shock fronts on the clumps and, from analysis of line widths, study the possible effects of turbulence and dynamics around the H~II region.

 We describe  observation details in \S 2 and present the results in \S 3. In \S 4 we present the calculations and the models that we applied to determine physical parameters of the observed regions. \S 5 contains  a discussion of how the H~II region has affected the cloud physical characteristics and the paper is summarized in \S 6.

\section{Target Selection and Observations}

\subsection{Investigated Region}

At a distance of  $1.09\pm 0.21$ kpc  \citep{Rus2007}, S175 is the closest source in our sample and we are able to resolve smaller structures within the molecular cloud associated with this H~II region.  The H~II region has been excited by a  B1.5V star (M $\sim$ 9.5 M$_\odot$, Holmgren et al. 1997) at  $\alpha$(J2000) = $ 00^{h}27^{m}17.1^{s}$
 and   $\delta$(J2000) = $ +64^{o}42'18.0''$.   Scaife et al. (2008) calculated a dust temperature of $T_d=27.9$ K for S175 by fitting a modified Planck spectrum to  IRAS 100 and 60 $\mu$m flux densities and, using optical recombination lines, also calculated an electron temperature of $T_e=7000\pm200$ K.  Fich and Rudolph (in prep.) recently calculated a density of 112 cm$^{-1}$  and a total mass of 0.4 M$_\odot$ for this ionized gas.  Wouterloot \& Habing (1985) also had reported a $T_{CO}$(1-0)=2.5 K and $T_{ex}$=5.7 K  toward the associated molecular cloud, but they had low spatial and frequency resolution.
 
\begin{figure}
\includegraphics[angle=0,scale=0.45]{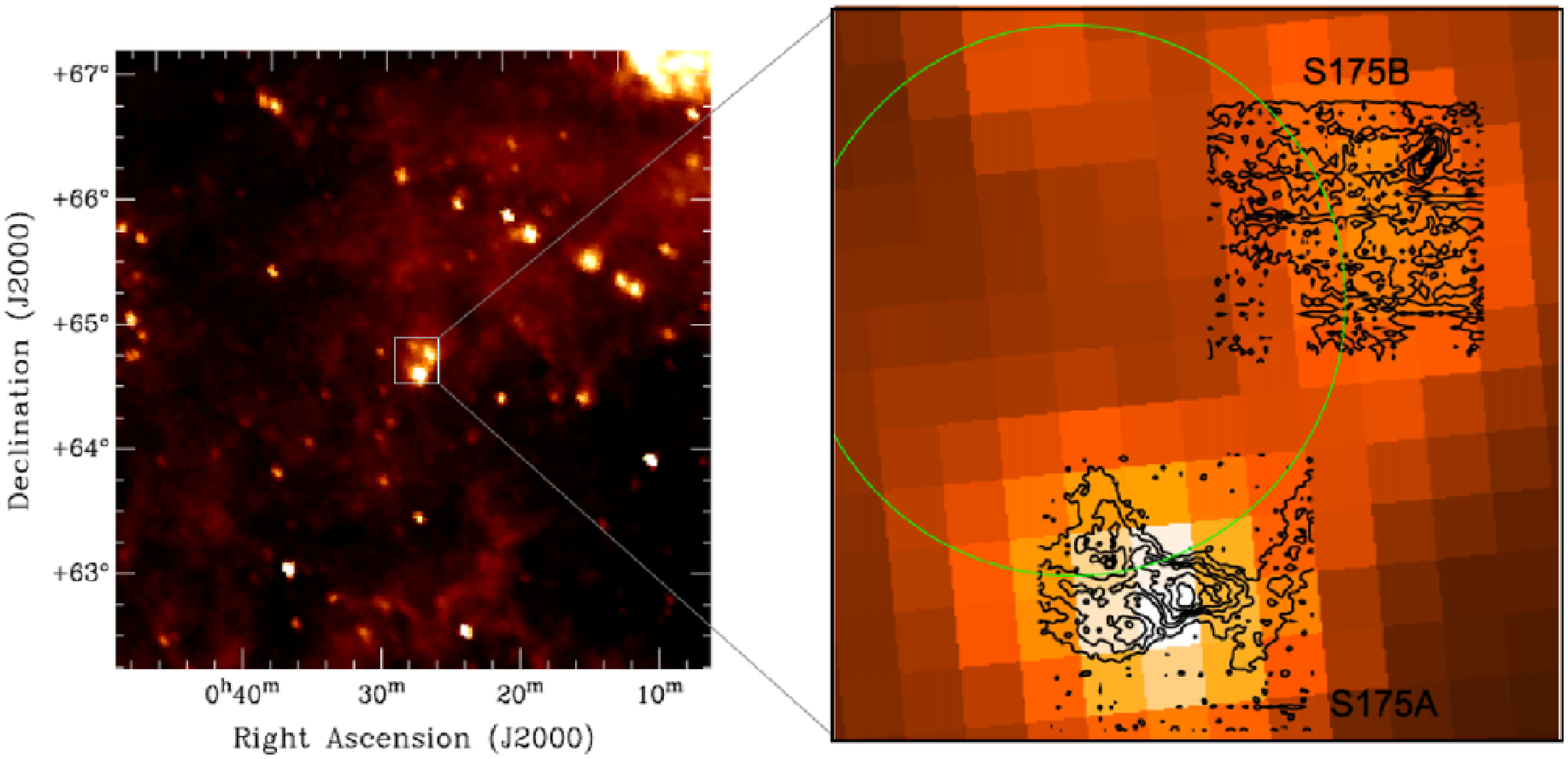}
\caption{ Position of S175 region in IRAS 12 $\mu$m map. Three condensations along a ring (shown in green) have been identified. The ring structure has also been observed at AMI 15.8 GHz \citep{ami2008}. Overlaid contours of integrated $^{12}$CO(2-1), shows the two $7'\times7'$ observed regions. S175A is adjacent to the H~II region while S175B lies at a distance of  $\sim 10'$ ($\sim$ 3 pc).
\label{iras}}

\end{figure}

Two components,  S175A and S175B, in the molecular cloud have previously  been identified in an IRAS  survey (of  H~II regions) by Chan \& Fich (1995). Both regions have the same $V_{LSR} \approx 50$ km s$^{-1}$ and are connected by a recently observed filament of molecular gas with the same velocity.  Therefore it is reasonable to assume that S175A and S175B lie at the same distance. Figure \ref{iras} shows the position of the cloud in an IRAS 12 micron map (left panel).    The components of the molecular cloud associated with the S175 H~II region cannot be resolved on the IRAS map but a recent study with the Arcminute Microkelvin Imager (AMI) \citep{ami2008} shows that S175A and S175B, observed at 15.8 GHz, sit on a ring of an extended emission (Figures 2 and 3 in their paper).  This ring, with a diameter of $~14'$ or $~$7 pc, is shown in green in Figure \ref{iras} (right panel).  A third condensation, which we label S175C, in the North-East of the boxed region can be seen but it  is not as intense as the two  mapped regions.  
 
 S175A is adjacent to the H~II region  and is likely  to be affected by the ionized gas while S175B, at a distance of $\sim 3$ pc from the visible edge of the H~II region (ten times the Stromgren  sphere radius),  is too far away to be affected. Therefore comparison of these two regions  may provide significant insight  into the effects of   the formation of a massive star (and the subsequent expansion of  ionized gas)  on the molecular cloud environment and on the physical properties that may affect future star formation.
  
\subsection{JCMT Observations} 

In August and November, 1998, we made JCMT observations in $^{12}$CO(2-1) of the two selected regions, S175A and S175B,  (Figure \ref{iras}).  We made a $^{12}$CO(2-1)  $7'\times7'$ map for each  of these regions, with the A3 heterodyne receiver. The beam size at this frequency is  $21''$.  Data was taken by driving the telescope in right ascension (sampling step of $7''$) with a  4 second exposure time at each point.   We made a mosaic of 12 sub-maps to complete each of the S175A and S175B maps. Eleven  of the sub-maps consist  of 5 rows and 59 columns, and one consists of 4 rows and 59 columns to complete 59$\times$59 pixel maps.  

The $^{12}$CO(2-1) maps were used to study the structure of the S175A and S175B molecular cloud and to search for the  $^{12}$CO(2-1) peaks within the identified clumps (see the next section for clump identification rules).  $^{12}$CO(2-1)  is optically thick and can not be used to detect the dense gas embedded  within the clumps. Therefore, the position of the brightest $^{12}$CO(2-1) emission in each clump  were observed in the optically thinner emission lines, $^{13}$CO(2-1) and CS(5-4).  We made pointed observations toward these positions in  August 2005, December 2007 and July 2008.   $^{13}$CO(2-1) was detected towards all positions but we failed to detect any CS(5-4) emission within S175A.   The peaks detected in S175B display lower temperatures; therefore, no CS(5-4) emission was expected and accordingly no CS pointed observations were made for S175B. 

We used a bandwidth of 267.5  MHz with 1713 frequency channels which corresponds to a velocity range of about 350 km s$^{-1}$ with a resolution of $\sim$0.2 km s$^{-1}$ for $^{12}$CO(2-1). 
The $^{13}$CO(2-1) and CS(5-4) pointed observations in S175A were made in frequency switching mode. We used a frequency switch of 8.3 MHz, a velocity resolution  of  $\sim$0.05 km s$^{-1}$ and a velocity range of  225 km s$^{-1}$. Frequency switching mode was not available while observing S175B and therefore we used a position switching mode to observe $^{13}$CO(2-1).   The system temperature, $T_{sys}$, was typically 300-600 K for $^{12}$CO(2-1)  mapping and 400-500 K for pointed observations. The typical noise level was  2 K for $^{12}$CO(2-1) and 0.3 K for $^{13}$CO and CS. We used SPECX and the Starlink SPLAT and GAIA packages to  reduce the data, make mosaics, remove baselines and fit Gaussian functions to determine  $\Delta$V, the  FWHM of the  observed line profiles.

\section{Structure in S175A and S175B}

We used the $^{12}$CO(2-1) maps to study the structure and morphology of the molecular
cloud associated with S175, and to identify  clumps  within it.   The  cloud consists of various clumps of gas that display one or more peaks of emission (typically towards the centre of the clump), see Figures \ref{S175Aclumps} and \ref{S175Bclumps}. We define any separated condensation as a distinct clump, if: 1) the brightest peak within the region has an antenna temperature larger than five times the rms of the background noise;  2)   the drop in  antenna temperature between two adjacent bright peaks  when moving  from one  to another  is  larger than the background noise; and, 3) the size of the condensation is  larger than the telescope  beam size ($21''$ or 3 pixels).   The edge of a clump is taken to be the boundary at which the antenna temperature drops to below half of the highest measured temperature within that clump. An ellipse that best fits to this boundary  is considered for clump size and integrated flux measurements. Note that this definition means that clumps can contain several peaks of emission and, in fact, almost half of the detected clumps contain more than one bright peak.

\subsection{S175A}

\begin{figure}

\center
\includegraphics[angle=0, scale=0.4]{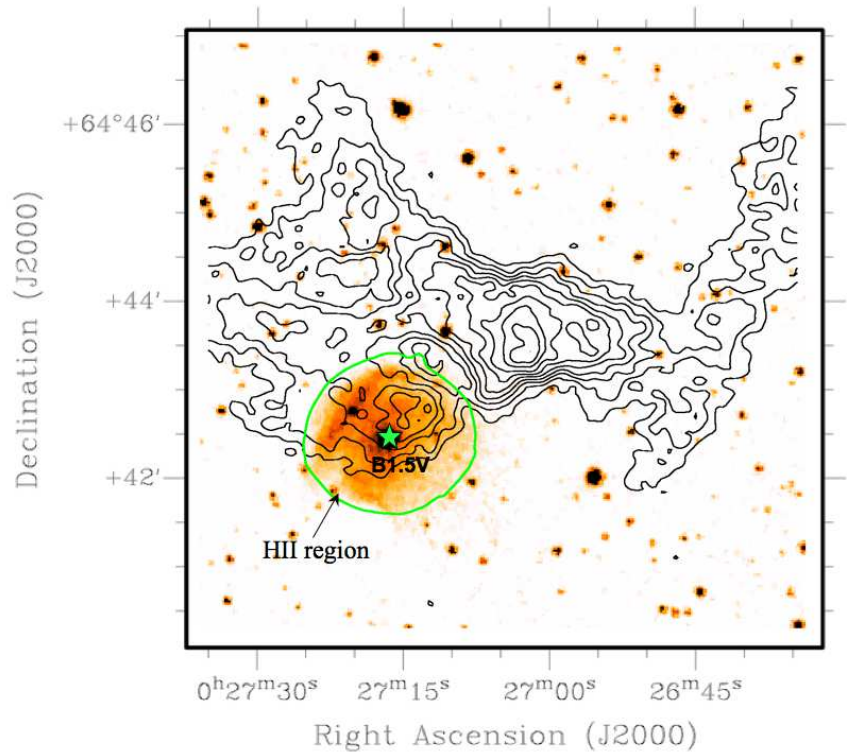}
\caption{S175A  integrated molecular gas contours  overlaid on an optical image from the  Digital Sky Survey of the H~II region S175. The visible borders of the H~II region are shown in green and the central bright source is an B1.5V star. The S175A molecular gas forms a wedge-shaped structure that overlays  the H~II region at South-East plus a filament that leads to the north-west. \label{S175A-1}}
\end{figure}

We find the gas velocity in S175A to range from -48 km s$^{-1}$  to
-52 km s$^{-1}$, which is consistent with previous studies that found  V$_{LSR}$(CO)= -49.6 $\pm$ 0.5 km s$^{-1}$ \citep{BlFi82} and -50.3 km s$^{-1}$ \citep{WoutHab85}. The $^{12}$CO(2-1) map (integrated between -48  km s$^{-1}$ and -52 km s$^{-1}$) overlaid on the optical image of the H~II region  is presented in Figure ~\ref{S175A-1} and  shows that S175A
takes the form of a wedge, which lies partially over the H~II region, and a filament that leads to the 
North-West.
 The brightest $^{12}$CO(2-1) emission peak lies at the centre of
 the region and at a velocity of -49.4 km s$^{-1}$. 

\begin{figure}
\hspace {-1.2cm}

\includegraphics[angle=0, scale=0.5]{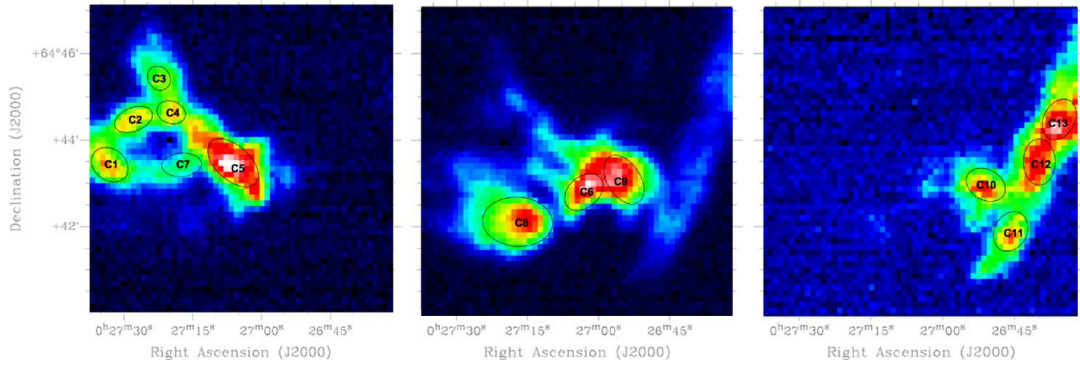}

\caption{S175A integrated over -48 km s$^{-1}$ to -49.5 km s$^{-1}$ (left), -49.5 km s$^{-1}$ to 
-50.5 km s$^{-1}$ (middle) and -50.5 km s$^{-1}$ to -52 km s$^{-1}$ (right ) . We detected 13 distinct 
clumps within this cloud. The position and other observed physical properties of each clump
are listed in Table \ref{tbl-1}. \label{S175Aclumps}}
\end{figure}

Figure ~\ref{S175A-1} also reveals S175A to have a clumpy structure.  We used the clump selection method described above to identify thirteen clumps within S175A, which are indicated in Figure \ref{S175Aclumps} by ellipses and listed in Table \ref{tbl-1}. 
 
S175A consists of a bright centre  containing clumps C5, C6 and C9. Flanking the central area to the East is a ring of clumps, of diameter $\approx$ $2'$,  consisting of C1-4 and C7 (Figure \ref{S175Aclumps} left panel).   We find no evidence for expansion of this ring; all the clumps display narrow single peaks at the same velocity (-49.5 $\pm$ 0.5 km s$^{-1}$) .  A chain of clumps consisting of C10-13 lies to the West of the centre (Figure \ref{S175Aclumps} right panel). The C8 clump, south of the ring, is the only clump in the same line of sight of the H~II region.  We are unable to determine whether C8 lies  in front of or behind the H~II region. 

 $^{12}$CO(2-1) is optically thick and can not be used to detect the dense gas within the clumps.
Therefore, we observed the brightest peak  within each clump using single point observations of
$^{13}$CO(2-1) and CS(5-4), in order to further probe the properties of the clumps in S175A.    The observed antenna temperatures, $T_a^*$,  at these points are listed in Table \ref{tbl-1}, and range from  9.6-29.1 K for $^{12}$CO(2-1)  and  2.26-17.34 K for  $^{13}$CO(2-1).  We failed to detect any  CS(5-4) emission within this cloud, indicating that the densities of the dense centres are less than  $\approx$ 10$^6$  cm$^{-3}$.
   
As noted above, S175A is situated very close to the H~II region and it could be expected to exhibit signs of turbulence, which could manifest as a large velocity dispersion between the clumps and/or as asymmetric line profiles broadened significantly beyond the thermal line width.   However, we find the spread in velocity of the brightest peaks to be small: -49.0 to -50.8  km s$^{-1}$ for $^{12}$CO(2-1) and -49.0 to -50.9 km s$^{-1}$ for $^{13}$CO(2-1).    Furthermore, both  $^{12}$CO  and $^{13}$CO  velocity profiles of the peaks are very narrow   ($ 0.78<\Delta V< 1.61$ km s$^{-1}$ for $^{12}$CO and    $ 0.47<\Delta V<1.08 $ km s$^{-1}$    for $^{13}$CO - see Table \ref{tbl-2})  and symmetric.  Comparing these line widths with calculated thermal line widths calculated from $T_{ex}$  found in \S\ref{TempTau} ($  0.51<\Delta V< 0.82$ km s$^{-1}$), and assuming  $T_{kin}=T_{ex}$, we may conclude that  S175A is quiescent region and not dynamically active.

\subsection{S175B}

\begin{figure}
\rotate
\hspace{-0.8cm}
\includegraphics[scale=0.4]{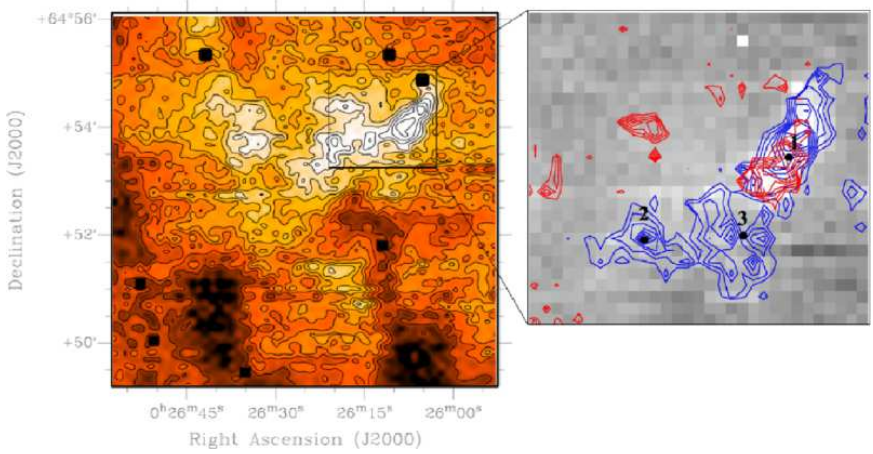}
\caption{S175B  $^{12}$CO(2-1) map integrated between -47 km s$^{-1}$ and -53 km s$^{-1}$. Three dense clumps within the box   show evidence of outflow  (notice the wide double peaked spectra for these clumps  in  Figure \ref{S175Bspecs}).   The red contours integrated over  -47 to -48 km s$^{-1}$ and blue contours over -53 to -54 km s$^{-1}$.  Our recent observations in $^{12}$CO(3-2) (Azimlu, M$^{\textrm c}$Coey \& Fich, in prep.) shows strong  signature of an outflow in C1.   The rest of the cloud, is  uniform and lies on extended diffuse gas with a temperature of $^{12}T_a^* \sim 5-7$ K.\label{S175B1}}
\end{figure}

\begin{figure}
\hspace{-1cm}
\includegraphics[angle=0, scale=0.6]{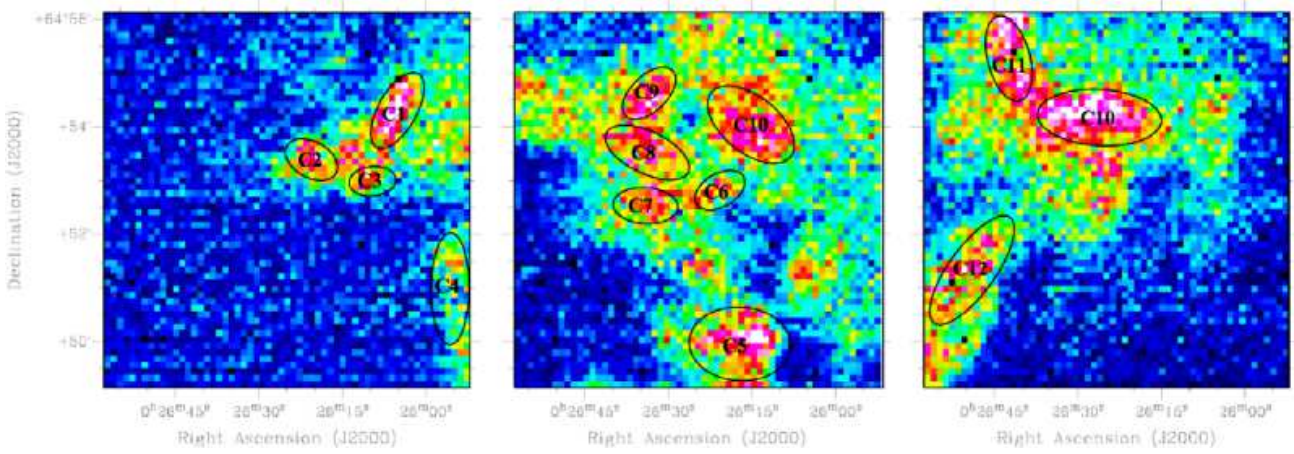}
\caption{S175B  $^{12}$CO(2-1)  map integrated over -47 km s$^{-1}$ to -49 km s$^{-1}$ (left), -49 km s$^{-1}$ to 
-51 km s$^{-1}$ (middle) and -51 km s$^{-1}$ to -53 km s$^{-1}$ (right ). It was more difficult to identify clumps in S175B because of extended gas with $^{12}T_a^{*}\sim 5-7 K$  We detected 12 distinct 
clumps within this cloud. The position and other observed physical properties of each clump
are listed in Table \ref{S175Bt1}. \label{S175Bclumps}}
\end{figure}

The S175B $^{12}$CO(2-1) map, which is integrated between -47 km s$^{-1}$ and -53 km s$^{-1}$,  is
 shown in Figure ~\ref{S175B1}.  In comparison with S175A,  S175B is more uniform with an extended CO emission averaging $\sim$ 5-7 K  covering more than 80 \% of the observed region (Figure \ref{S175B1}, left panel).  Furthermore, a consistent background  noise of $\sim$ 2 K exists through the mapped region.  
 
Clump identification was therefore more difficult in S175B and was restricted to a region at the north-west that showed a greater degree of dynamical activity than in the rest of the region.  A particularly dynamically active area is highlighted in the right panel (enlarged square) of Fig. \ref{S175B1}.   Twelve distinct clumps have been resolved within S175B (Figure \ref{S175Bclumps}) and are indicated by ellipses, with corresponding positions listed in Table \ref{S175Bt1}.    Most of the clumps are located in the northern part of the cloud and only three clumps (C4, C5 and C12) have been identified in the southern half.  We are unable to resolve these three clumps any further because the southern half of the cloud is especially uniform and the clumps have irregular shapes with  edges that are hard to distinguish from the extended  5-7 K  CO emission.

As for S175A,  single point observations in $^{13}$CO(2-1) were made at the peaks within the clumps (C12 is yet to be observed in  $^{13}$CO).  The observed physical parameters of each of these positions are listed in Table \ref{S175Bt1}.  We note here that the observed T$_a^*$ are generally lower in S175B than S175A ($^{12}T_a^{*}$=9.3-16.8 K and $^{13}T_a^{*}$=2.7 K to 6.3 K), while S175B shows a slightly wider velocity range at the peaks, with -47.8 km s$^{-1}$ $<$ V($^{12}$CO) $<$ -52.2 km s$^{-1}$  and -47.7 km s$^{-1}$ $<$V($^{13}$CO)  $<$ -52.0 km s$^{-1}$.  

In contrast to the narrow and uniform  $^{12}$CO(2-1) line profiles ($\Delta V \sim1$ km s$^{-1}$) of the extended gas, the peaks within each clump have wide line profiles
($ 1.04<\Delta V< 8.68$ km s$^{-1}$ for $^{12}$CO and  $0.59<\Delta V<2.11 $ km s$^{-1}$     for $^{13}$CO - see Table \ref{S175Bt2}).  With the exception of C5, which lies in a somewhat isolated position at the centre in the southern half of the region, the peak position $^{12}$CO line profiles are asymmetric. S175B is clearly more dynamically active than S175A.
Comparing the emission line widths with thermal line widths, ($  0.51<\Delta V< 0.65$ km s$^{-1}$) confirms that the observed line widths are significantly larger than the thermal velocity dispersion.

\begin{figure}
\epsscale{0.8}
\plottwo{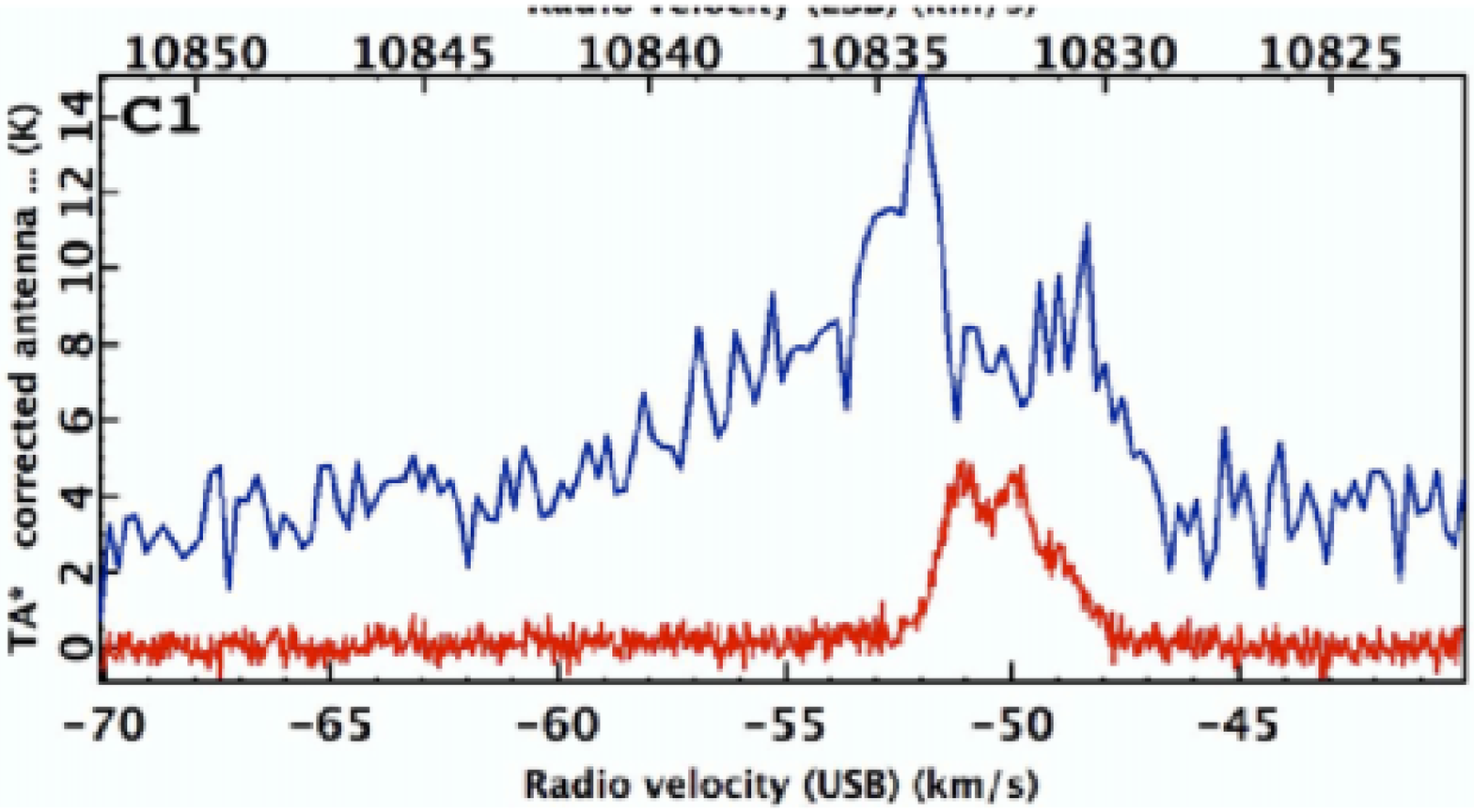}{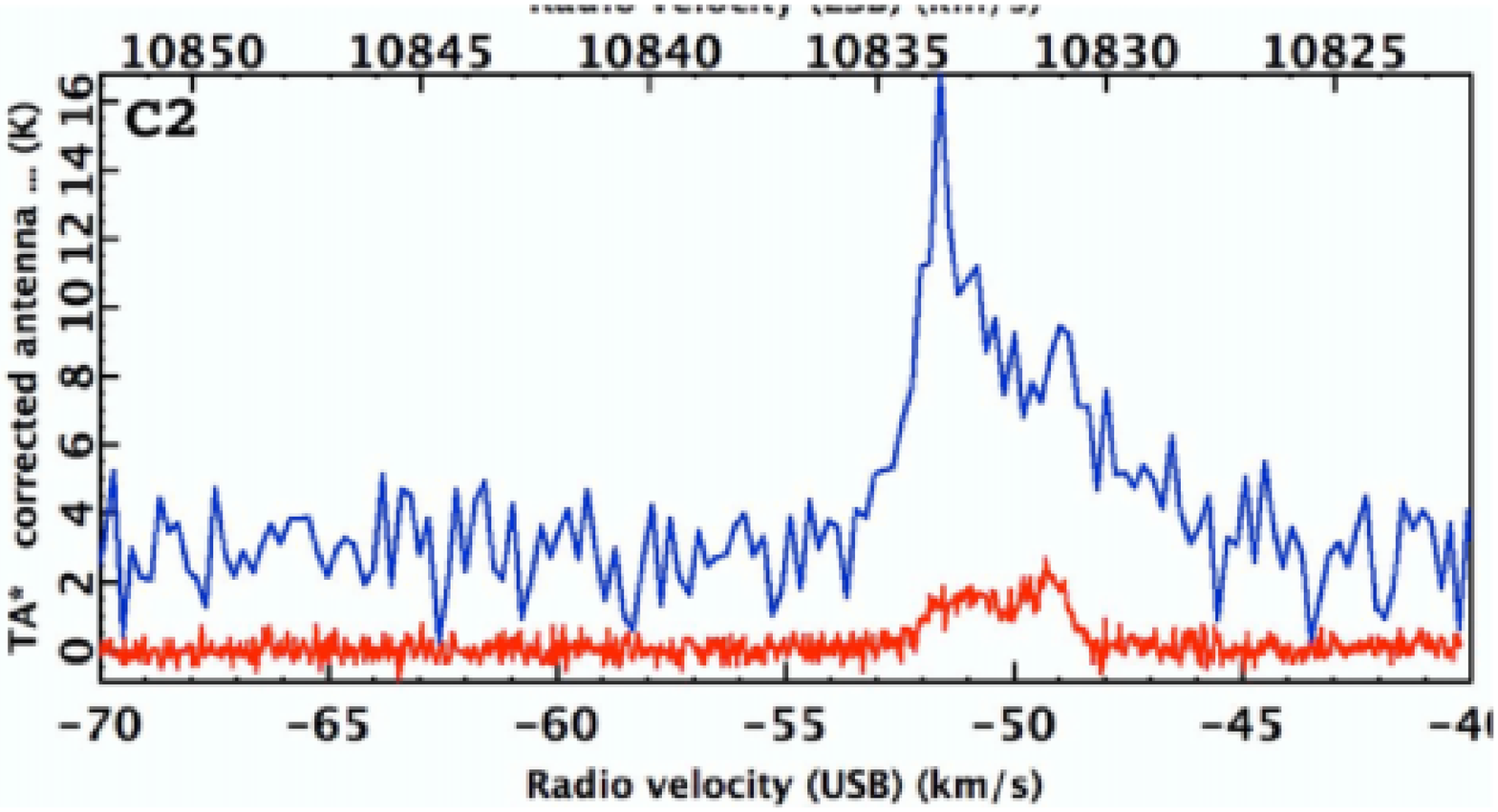}\\
\plottwo{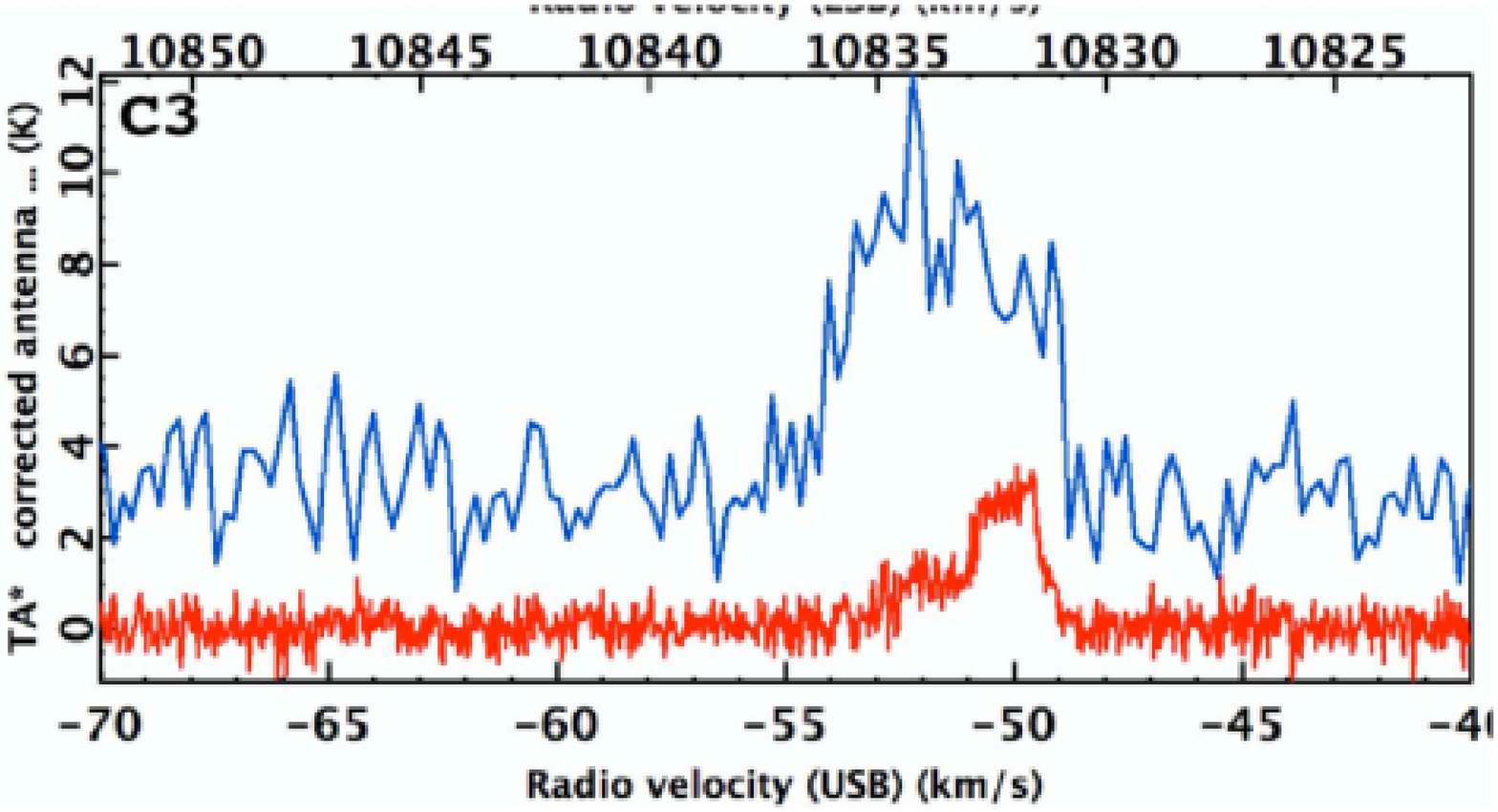}{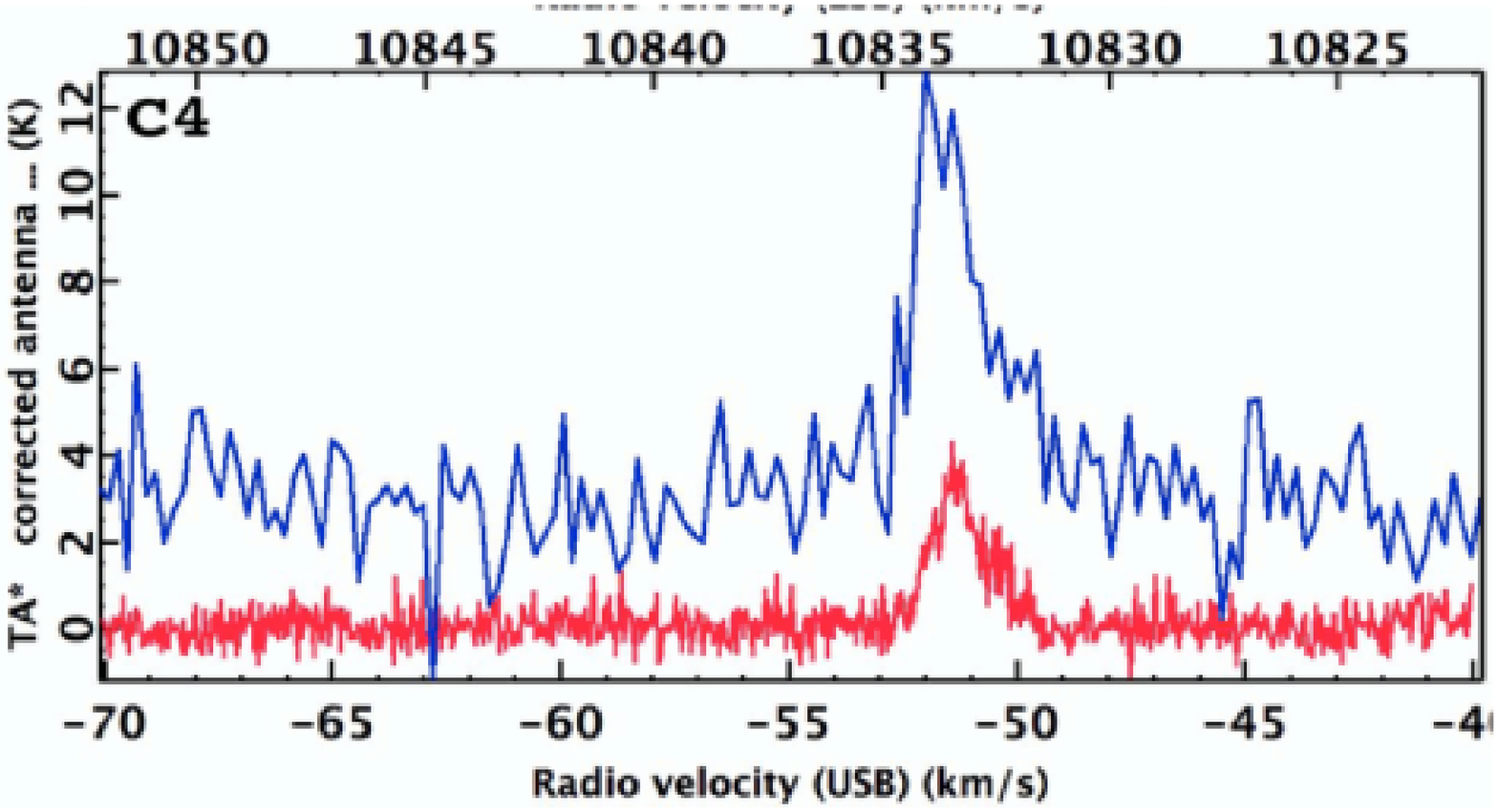}\\
\plottwo{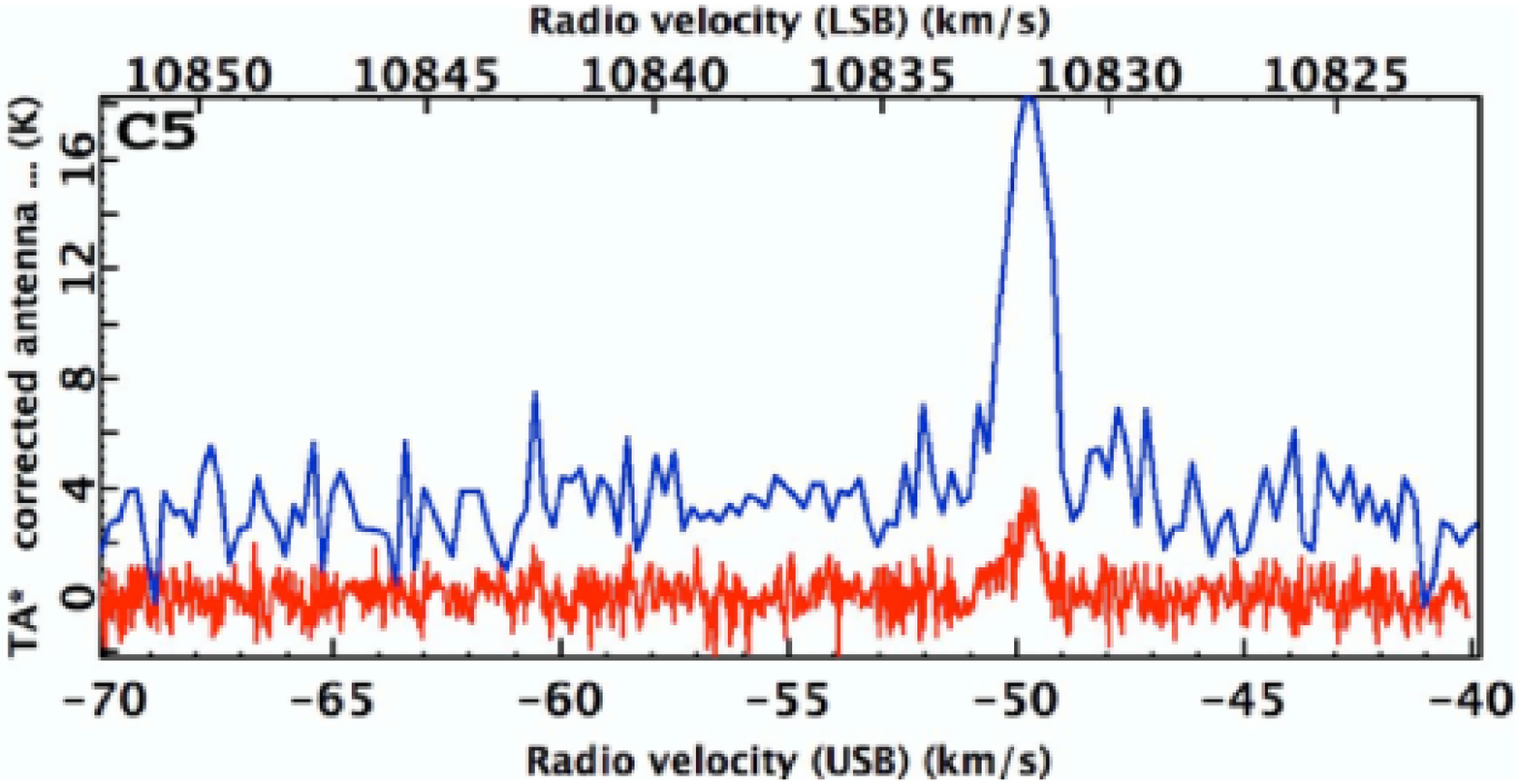}{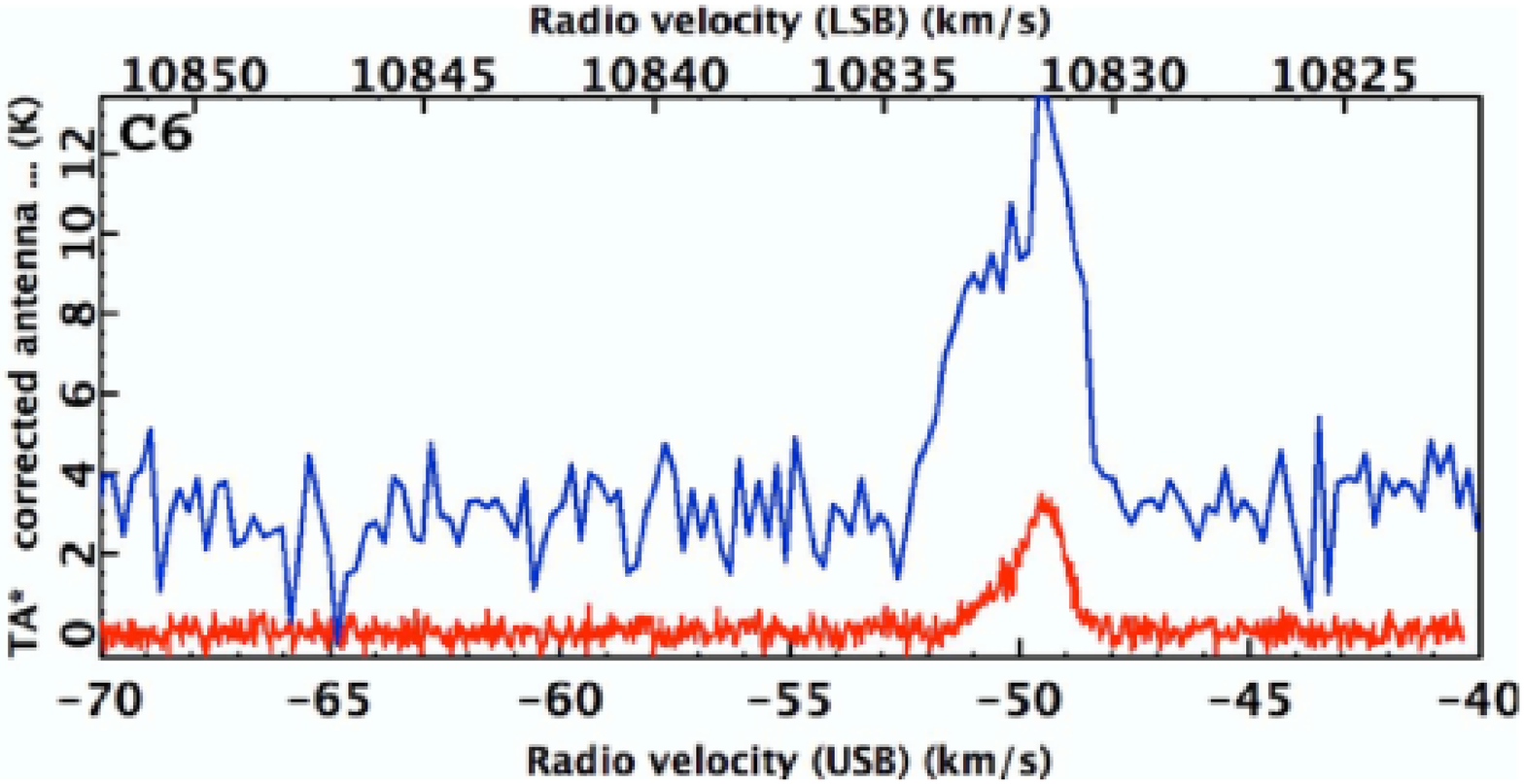}\\
\plottwo{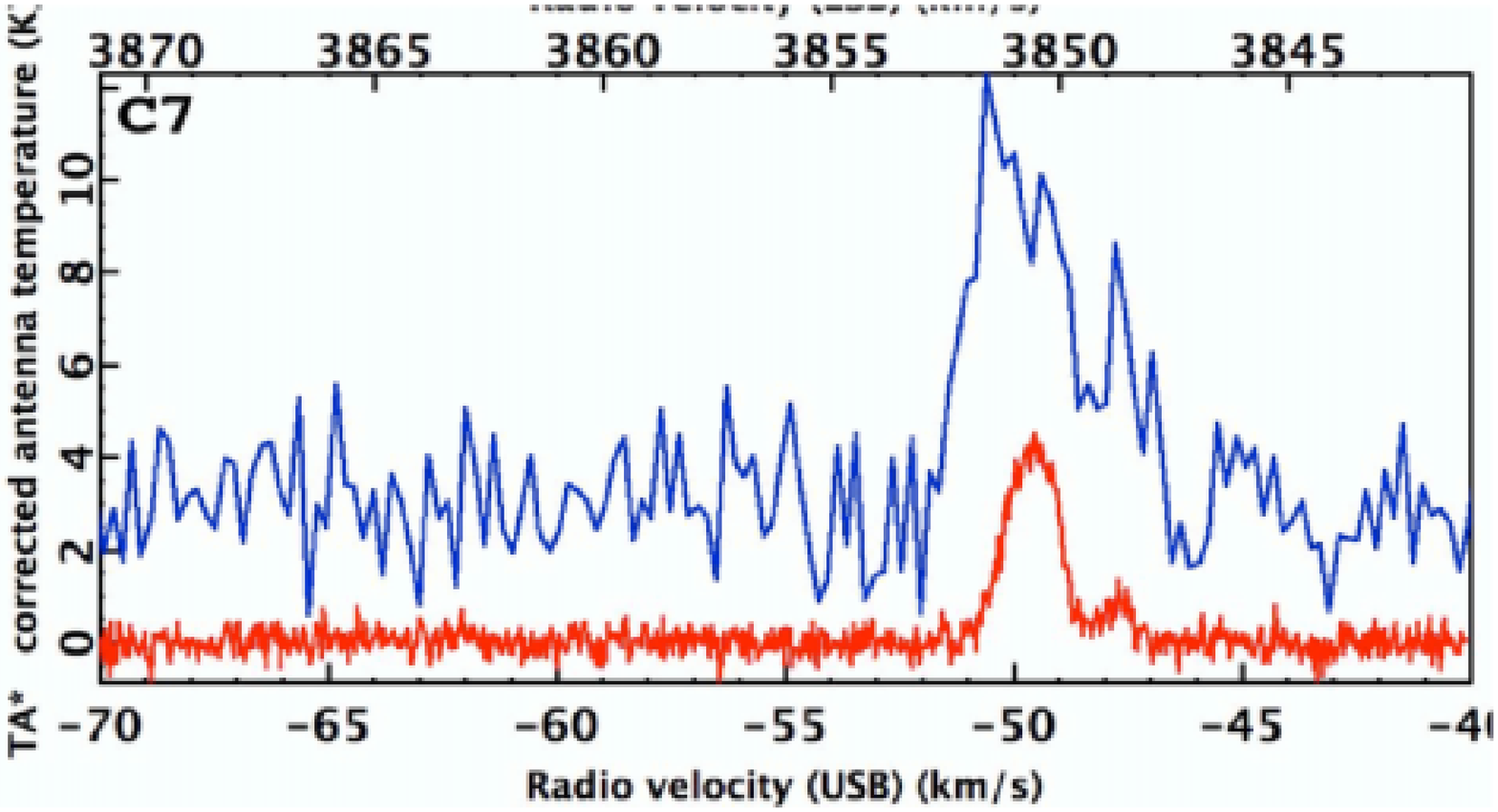}{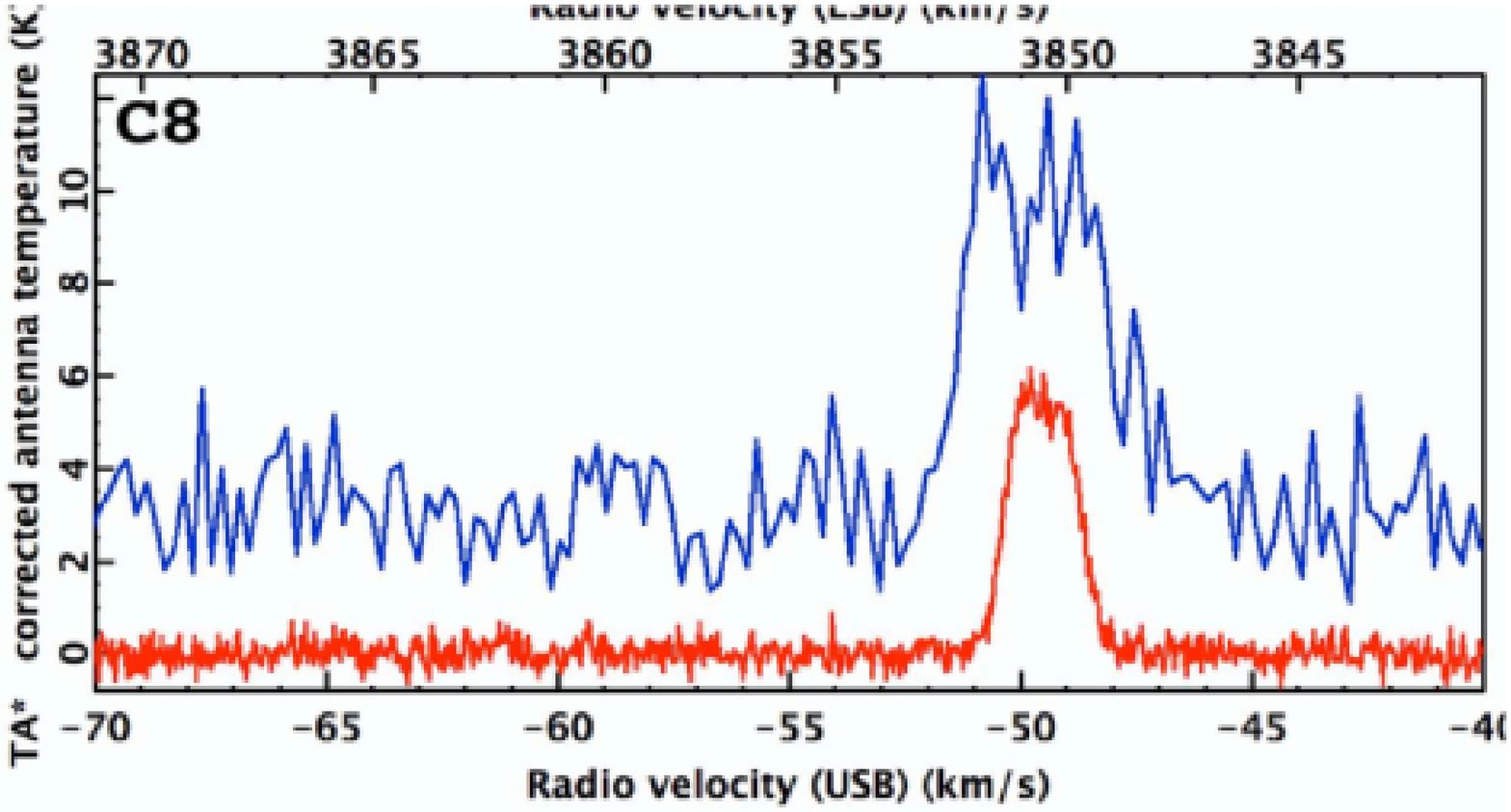}\\
\plottwo{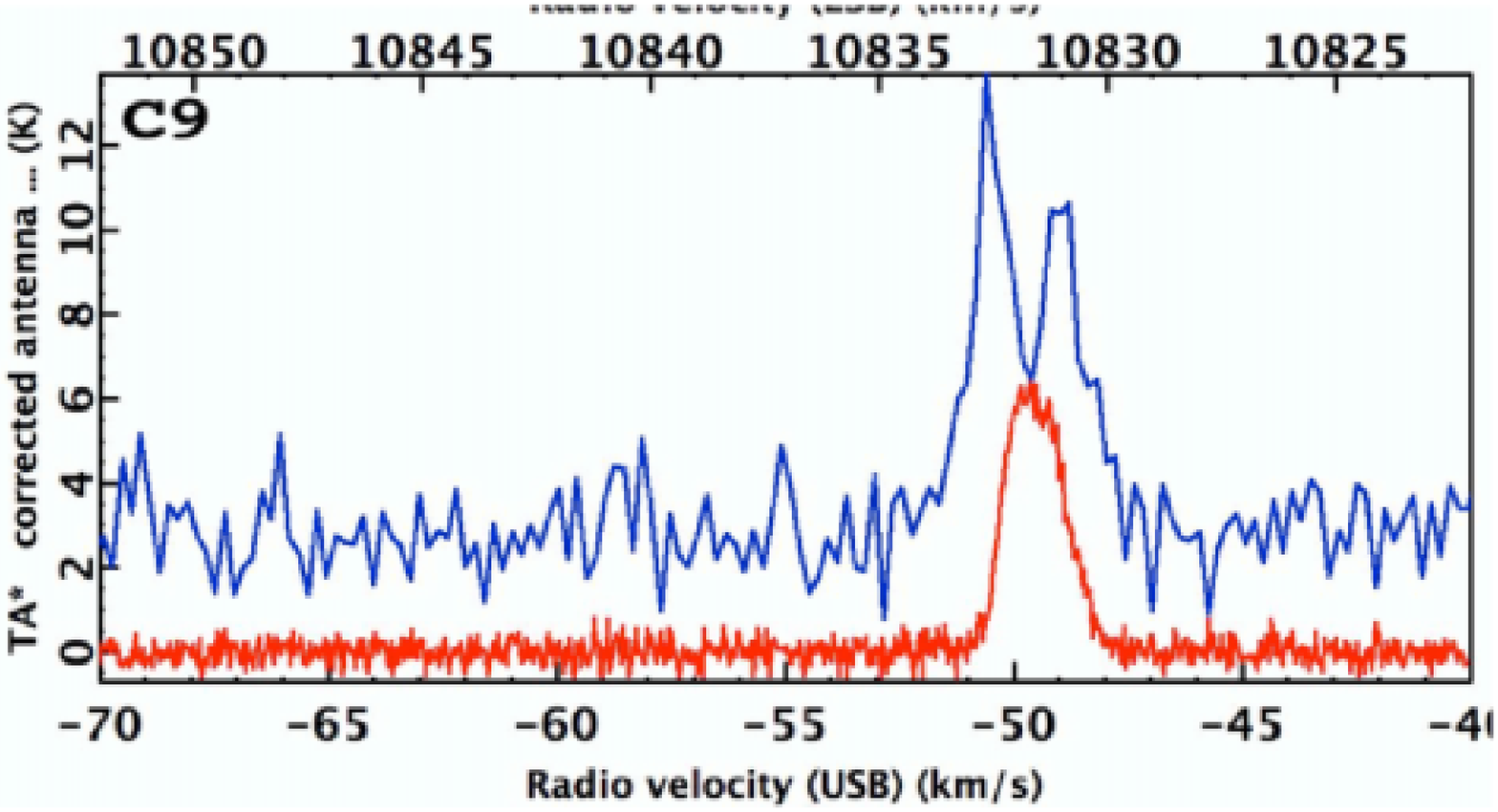}{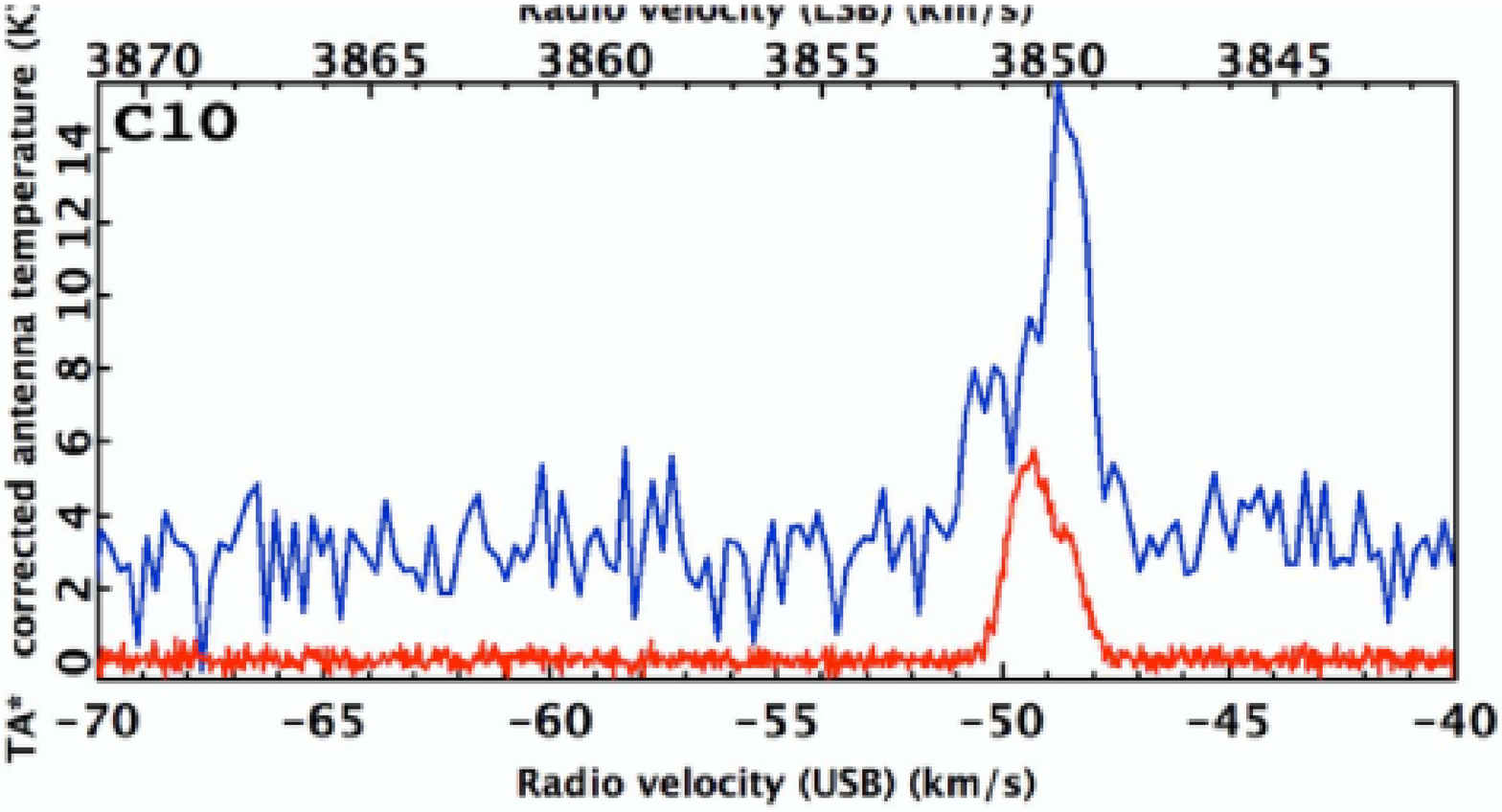}\\
\plottwo{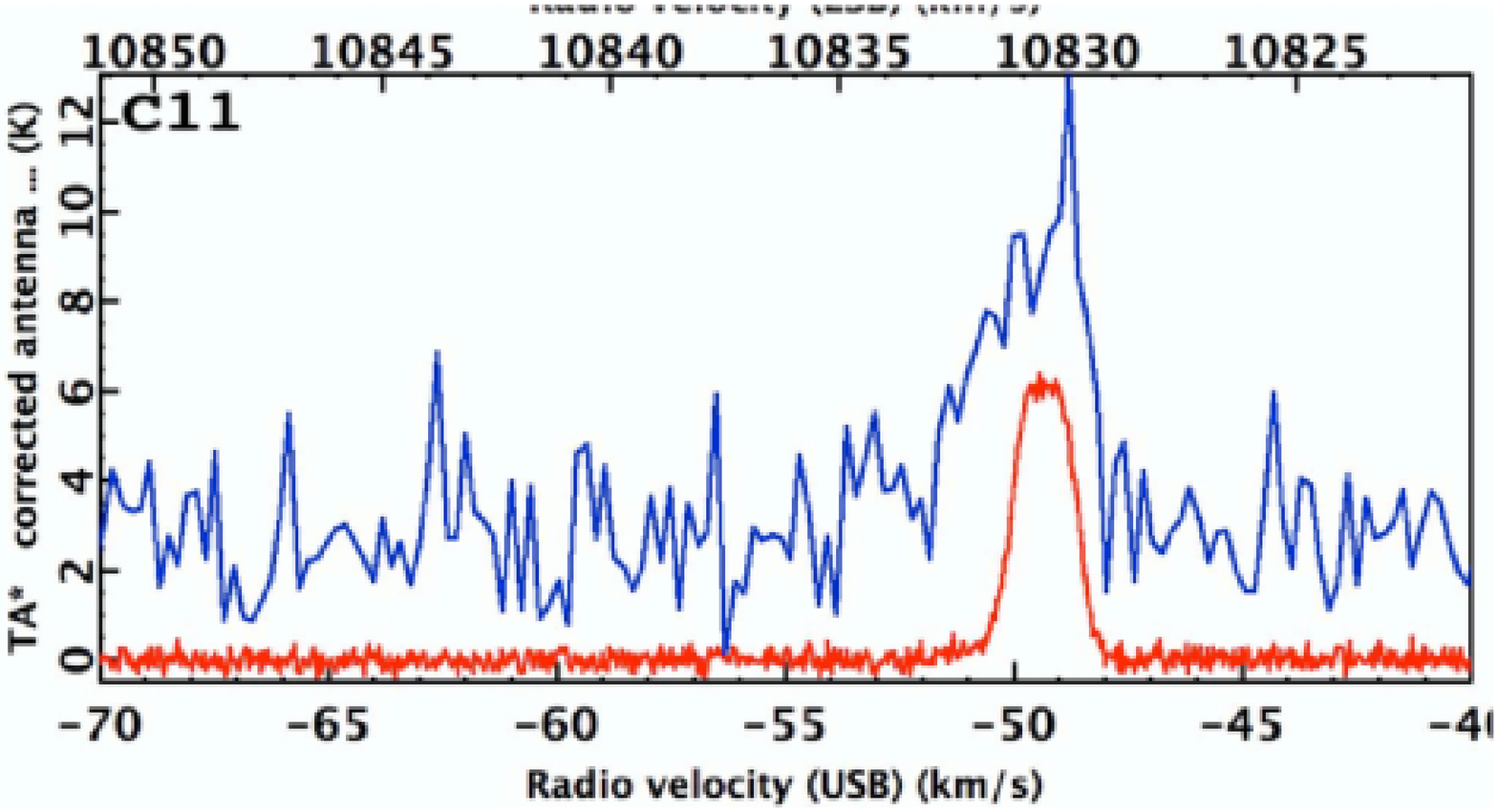}{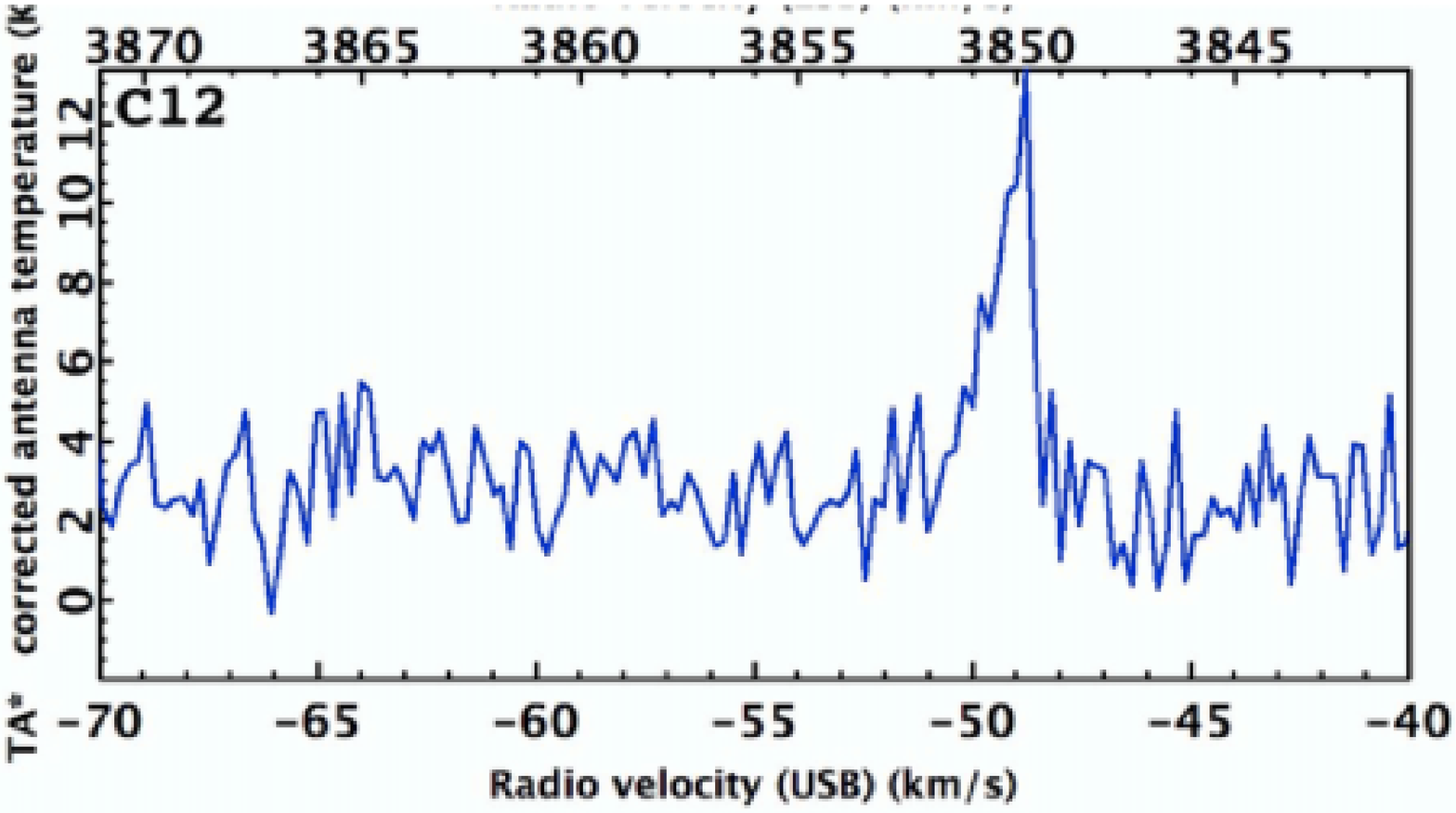}

\caption{$^{12}$CO(2-1) (blue spectra) and $^{13}$CO(2-1) (red  spectra). For clarity, the $^{12}$CO(2-1) profile has been shifted upwards by 3 K. Note the  broad, distorted lines in C1-C3 which are representative of an outflow. Strong self absorption is seen in C8 and C9. C12 is yet to be observed in $^{13}$CO(2-1) (see Figure 7 for C11 and C12 ). \label{S175Bspecs}}
\end{figure}

Furthermore, we see multiply-peaked profiles  in $^{12}$CO(2-1), which are identified by letters in Table \ref{S175Bt2},  in C1-C3, C7 and C9  (compare the line widths and profile shapes  in Figure \ref{S175Bspecs}). The multiple peak structure could be  a result of self-absorption or multiple cores at different velocities along the line of sight.  In $^{13}$CO(2-1), we find that C1-3 and C7-11 appear to be either saturated or self-absorbed.   Signatures of self absorption and saturation  in  $^{13}$CO can be an indication of warm gas inside the cores.  Alternatively, multiple peaks in $^{13}$CO could support the possibility of multiple cores. Our recent observations in $^{13}$CO(3-2) (Azimlu, M$^{\textrm c}$Coey \& Fich, in prep.) at C1-C3 shows strong signature of an outflow located at C1, which could explain multiple peaked profiles for these peaks.

\section{Physical Parameters derived from CO Observation}

 $^{12}$CO(2-1) maps have been used to study the morphology of the cloud, and to identify the position of clumps and their size. These observations also are used to determine the  brightest peaks within each clump and measure the physical parameters of each peak, such as antenna temperature, line width and velocity profiles. Using these parameters we can calculate the mass of the clumps assuming Virial equilibrium conditions or calculate the velocity integrated (or ``X-factor'') mass, using  a known ``X-factor''   (an empirical ratio of H$_2$ column density to the velocity integrated $^{12}$CO(1-0) emission) to calculate  the  H$_2$ column density. 
 
 $^{12}$CO is optically thick and not suitable to measure the properties of the densest  gas; therefore, we also calculated mass using observations of the brightest peak within each clump in  $^{13}$CO(2-1). We measure the  $^{13}$CO(2-1) antenna temperature and line width directly from the observed spectrum. Then, assuming  Local Thermodynamic Equilibrium (LTE), we determine the excitation temperature, opacity and gas column density for the  observed points.  Assuming that clumps are uniformly spherical, we calculate the average volume density, LTE mass, and the Jeans Mass for each clump.

\subsection{Temperature and Opacity}
\label{TempTau}

The corrected antenna temperature, $T_a^{*}$,  and line widths for both $^{12}$CO(2-1) and
$^{13}$CO(2-1) emission lines were directly measured from spectra, after baseline subtraction.  For multiply peaked profiles, we determine a separate line width for each peak only if the separation is greater than 1 km s$^{-1}$ (five times of  the channel resolution of 0.2 km s$^{-1}$).   The brightest $^{12}$CO peak within a multiply peaked spectrum is used in calculation.

$T_b$, the received brightness temperature,  is related to the directly observed antenna temperature through the beam efficiency and beam filling factor; $T_a^*= T_b\eta_{mb}f_{BEAM}$.  The beam efficiency, $\eta_{mb}$,  is 0.69 at the JCMT for the observed frequencies, and $ f_{BEAM}$ is the fraction of the telescope beam filled by the source emission.  Although our sources are larger than the beam size, we expect that there will be sub-structure due to the power-law nature of the ISM.   However, we have no means to measure this and we take  $f_{BEAM}=1$ in order to calculate T$_b$.  Strictly speaking, therefore we find a lower limit to T$_b$. 

The opacity, $\tau$,  and  excitation temperature, T$_{ex}$,  can be derived by comparing the  $^{12}$CO(2-1) and  $^{13}$CO(2-1) brightness temperatures:

\begin{equation}
T_{b}= T_0\left [ f(T_{ex}) -
f(T_{bg})\right ]
\times [1 - exp(-\tau)]
\label{TbExp}
\end{equation}

\begin{equation}
\textrm{and} ~ ~~  
f(T)=\frac{1}{exp(T_0/T)-1},
\end{equation}

 where T$_{bg}$ is the microwave background temperature (2.73 K) and $T_0= h\nu/k$, with $\nu_{12}=230.538$  GHz for $^{12}$CO(2-1) and $\nu_{13}=220.399$  GHz  for $^{13}$CO(2-1).   We consider a Local Thermodynamic Equilibrium (LTE)  model in which  $^{12}$CO(2-1) and   $^{13}$CO(2-1) have the same
$T_{ex}$  and $\tau_{13}\ll \tau_{12}$. The abundance ratio of two isotopic species is also required in order to derive $T_{ex}$ and
$\tau$ from expression \ref{TbExp}. This ratio varies  from 24 at the centre of the Galaxy to a terrestrial value of 89. We adopt the commonly used average
value of 62$\pm$4 derived by Langer \& Penzias (1993) for molecular clouds  in the solar neighbourhood.

T$_{ex}$  can  be calculated from the optically thick $^{12}$CO line (e.g. Pineda et al. 2008). We can then solve equation (1) to obtain T$_{ex}$,

\begin{equation}
T_{ex}(^{12}CO)= \frac{T_0^{12}} {\ln \left [ 1 + \frac{T_0^{12}}{T_{b}(^{12}CO) + T_0^{12}f(T_{bg})}
 \right ]} \hspace{0.1cm},
\end{equation}
and then the opacity, using $^{13}$CO:
\begin{equation}
\tau_{13}= -\ln \left [ 1- \frac{T_{b}(^{13}CO)}{T_0^{13}} \left \{ \left [ exp(
\frac{T_0^{13}}{T_{ex}} ) -1 \right ]^{-1} - f(T_{bg})\right \}^{-1} \right  ].
\end{equation}
\\
To derive these equations  it has been assumed that  $\tau_{13}\ll 1 $ and $  \tau_{12} \gg 1$.  
The calculated $T_{ex}$ varies from 15.9 K to  34.5 K for S175A and from 14.2 K to 22.1 K for S175B. Observed and calculated parameters, such as T$_a^*$, T$_{ex}$,  $\Delta V_{12}$ and $\Delta V_{13}$,  $\tau_{13}$ and $\tau_{12}=62\times \tau_{13}$, and the corresponding hydrogen column density are listed in  Tables \ref{tbl-2} and \ref{S175Bt2} for S175A and S175B. 

The largest opacities in the two regions are found to be $\tau_{13}=0.90$ (C5 in S175A)  and $\tau_{13}=1.01$ (C8  in S175B).   C5, which is the brightest peak in S175A,  lies in the condensed central part of the region and could therefore be expected to have the highest opacity, and consequently the largest column density. 
However these two calculated values might be very inaccurate because to derive these results we have assumed that $\tau_{13}$ is much smaller than  unity which is inconsistent with the results.   We need to observe optically thinner emission lines such as C$^{18}$O to be able to estimate an accurate column  density for  these dense clumps.

\subsection{Mass Estimation}

In this Section, we describe how individual emission lines of  $^{12}$CO and $^{13}$CO are used to estimate the  mass of individual clumps in S175A and S175B by several means. We use line widths of  $^{12}$CO and $^{13}$CO in order  to calculate the Virial mass, we have also calculated  the  $^{12}$CO velocity integrated mass (or ``X-factor'' mass), and used an LTE model to determine the  $^{13}$CO column density.   We also calculate the Jeans Mass for each clump.
In the following we describe each method individually and give the range of masses found for S175A  and S175B.  The masses determined by each method for each clump can be found in Tables \ref{tbl-3}    and \ref{S175Bt3}.

\subsubsection{Virial Mass}
\label{VMass}

According to the Virial theorem for a stable, self-gravitating, spherical distribution of mass, the kinetic energy must equal the half of the potential energy. Assuming that our clumps are in Virial equilibrium, the Virial mass is calculated  from

\begin{equation}
M_{vir}(M_{\odot})=126\times R_e(pc) (\Delta V)^2(km s^{-1}) ,
 \end{equation}   
\\
which assumes a spherical distribution with  density  proportional to  r$^{-2}$ \citep{McLaren88}. 
If the calculated Virial mass is larger than the mass measured by other techniques then the object has too much kinetic energy and is not stable.
Most of the
clumps have an irregular shape and an appropriate value for the radius can not be taken from
the  $^{12}$CO maps. Instead, we calculate the area of a clump from the $^{12}$CO maps and assumed  the effective radius  of  $R_e = \sqrt{Area/\pi}$.  We do not have any measurements 
of the clump size from the  $^{13}$CO(2-1) pointed observations; however, it is a reasonable assumption that  the $^{12}$CO and $^{13}$CO lines are emitted from
the same  volume of gas (Rohlf \& Wilson 2004, LTE assumption - see \S\ref{LTE}). Therefore, we use the same  measured  radius from
 $^{12}$CO maps for both lines.

There are uncertainties in measurement of the parameters in this equation which carries through to the mass estimation. For example, our size estimation relies on distance measurement that has  an estimated  uncertainty  of $\sim$ 20$\%$. In addition, each CO isotopomer presents different line widths; $\Delta V_{12} $ is observed to be on average larger than  $\Delta V_{13} $ by a factor  of 1.7.   $^{12}$CO(2-1) is optically thick and most  of the observed lines are saturated and may show self-absorption; hence, we cannot measure the correct $\Delta V$. 
 
In some cases the difference is much larger due to presence of multiple peaks, which are not resolvable in  $^{12}$CO. For example S175B-C1 presents  a broadened  spectrum  in $^{12}$CO with  $\Delta V_{12} $=8.68 km s$^{-1}$. This spectrum is divided to three peaks in  $^{13}$CO with the main peak  $\Delta V_{13} $=1.44 km s$^{-1}$ which is six times smaller than $\Delta V_{12}$.  As a result,  the calculated   Virial mass using  $^{12}$CO(2-1) line width,  $M_{vir}$($^{12}$CO),  is overestimated by a factor of $\sim36$.  $M_{vir}$($^{13}$CO), which is calculated using $\Delta V_{13}$, then is the best estimation if the cloud is in Virial equilibrium. However, we note that self-absorption can be seen in a few cases for the optically thinner line of $^{13}$CO.  $M_{vir}$($^{13}$CO), ranges between 3.5-24.9 $M_\odot$  for S175A, and between 17-62 M$_\odot$ for S175B. In the case of self-absorbed  or multiple peaked lines, we are measuring too large a line width and consequently overestimating the Virial masses. 

\subsubsection{Velocity Integrated  Mass}

In the case of the optically thick $^{12}$CO emission we can calculate the column density from  the empirical relation $N(H_2) = X \times \int T_{mb}(^{12}CO)dv$, with $T_{mb}= T_a^*/\eta_{mb}$. For the JCMT,  $\eta_{mb}= 0.69$ is the beam efficiency at the observed frequency range.
$X$ is an empirical  factor which gives the ratio of 
 the H$_2$ column density  to the integrated intensity of $^{12}$CO(1-0). Both theory \citep{Bell06} and
 observation \citep{Pin08} indicate that the ``X-factor'' is sensitive to variations in physical parameters,  such as density, cosmic ionization rate, cloud age, metallicity  and turbulence. The ``X-factor'' also depends on the cloud structure  and varies from region to region. Studies (e.g. Pineda  2008,  and references therein) have shown that the $X$ value is  roughly constant for the observed  Galactic molecular clouds, and is currently estimated  at $X\simeq 1.9 \times 10^{20}$ cm$^{-2}$(K km s$^{-1})^{-1}$
for $^{12}$CO(1-0)   \citep{StMatx96}.  We use  $^{12}$CO(2-1) and so need to  correct  this factor.    Using a $^{12}$CO(1-0) point observation \citep{BlFi82} and the $^{12}$CO(2-1) smoothed  to the $^{12}$CO(1-0) beam size (2.3$'$), we calculated a ratio of $^{12}$CO(2-1)/$^{12}$CO(1-0) = 0.8, which is identical to  the value found by Brand \& Wouterloot (1998) in a larger sample.
 
$\int T_{mb}(^{12}CO)dv$ is derived by integrating over all pixels within an ellipse
best-fitted to the half highest peak  contour (Figures \ref{S175Aclumps} and \ref{S175Bclumps}). About  90\% of the integrated flux lies within this ellipse for  most of the individual distinct clumps. Adjacent or combined clumps  (C5, C6 and C9 within S175A and C10-C11 within S175B) have overlaps and  the common area between two fitted ellipses  has been integrated for both. For a few extended bright  clumps (C5, C6, C8 and C9 in S175A and C5 and C10 in S175B) we have considered the ellipse fitted to the
$3\sigma$ ($\sim $5 K) contour to define the edges of the clump which contains the $\sim$90\% of the integrated flux. 

The integrated mass is  determined by:

\begin{equation}
M_{int}^{12}=  D^2 m_H \mu \times X \int
 T_{mb}(^{12}CO)dvdA \hspace{0.1cm},
\end{equation}

\noindent where  D is the distance, $m_H$ is the mass of a hydrogen atom, and $\mu$= 2.33 is the mean molecular weight  for $H_2$ with a 25\% mass fraction of Helium.  
The calculated $M_{int}$ ranges from 4.1 to 19.4 M$_\odot$ for clumps in S175A,  and from 8.9 to 31.4 M$_\odot$ for clumps in S175B.

\subsubsection{LTE Mass}
\label{LTE}
The column density can also be calculated from the optically thin $^{13}$CO(2-1)
line under the assumption of LTE \citep{RW}:

\begin{equation}
N(^{13}CO) = 1.5 \times 10^{14} \frac{T_{ex} exp(T_{0( \nu_{10})}/T_{ex}) 
 \int
\tau^{13}(v)dv}{1 - exp ( -T_{0(\nu_{21})}/ T_{ex})}.
\end{equation}

\noindent where $ T_{0( \nu_{10})} = h\nu/k $  with $\nu_{10}=110.201$ GHz for $^{13}$CO(1-0) and  $\nu_{21}=220.399$ GHz for $^{13}$CO(2-1). When using the LTE model,  we assume that   $T^{12}_{ex}$=$T^{13}_{ex}$ and  that the excitation temperature is the same over the entire cloud. For optically thin lines, integrals involving $\tau(v)$  can be approximated by integrated line intensity

\begin{equation}
T_{ex} \int \tau(v)dv \simeq \frac{\tau}{1 - exp (-\tau)} \int
 T_{mb}(v)dv .
\end{equation}
\\
Assuming $N(H)\simeq 10^{6} \times  N(^{13}CO)$ \cite{Pin08}, the clump mass using the $^{13}$CO column density is then calculated as:

\begin{equation}
M_{LTE}= \frac{\pi}{3} R_{e}^2 m_H \mu N(H) .
\end{equation}
\\
$M_{LTE}$ ranges from   0.6 to 14.8 M$_\odot$ for S175A clumps and from 1.9 to 13.1 M$_\odot$ for S175B clumps.

\subsubsection{Jeans mass}

Assuming that the kinetic temperature, $T_K$, of the gas is equal to $T_{ex}$, the Jeans mass is given by

\begin{equation}
M_J = 1.0 M_{\odot} \left ( \frac{T_K}{10 K}\right)^{3/2} \left (
\frac{n_{H_2}}{10^4 cm^{-3}}\right)^{-1/2} .
\end{equation}

\noindent 
Only thermal pressure is considered in the calculation of the Jeans mass,  while some agent, such as turbulence and/or internal dynamics, may act to support the clump against gravitational collapse.   If that is the case then $M_{Vir}$ and $M_{int}$ will be larger than $M_{Jeans}$.  The clumps in S175A have a slightly larger $M_{Jeans}$ than those in S175B (4.3-12.7  M$_{\odot}$ versus 2.8-10.2  M$_{\odot}$).   The Jeans masses for each region have been listed in Tables \ref{tbl-3} and \ref{S175Bt3}.

\section{Discussion}  

The primary goal of this study is to investigate how the physical properties of  this molecular cloud have been affected by the the S175 H~II region. We have selected two different areas: S175A, adjacent to the H~II region, and S175B, which is distant enough that it is unlikely to be affected by the ionized gas around the exciting star. We have measured and calculated different physical parameters for two regions and will compare them in this section and discuss how they may have been affected by external sources of turbulence such as the S175 H~II region. 

\subsection{Comparison of the properties S175A and S175B}

\paragraph{Structure:}
The observed regions show significantly different structures. S175A, which appears to be affected by ionized gas,  has an elongated  structure and is fragmented into clearly identified distinct clumps, while S175B is very uniform and the clumps within it are less easily identified due to an underlying extended $^{12}$CO emission. The extended gas in S175B displays uniformly narrow lines ($\Delta V \sim 1$ km s$^{-1}$)  of  $^{12}T_a^{*}\sim5-7$ K.

\paragraph{Temperature:}
S175A  contains clumps with higher gas temperatures:  $^{12}T_a^{*}$  and $^{13}T_a^{*}$ range between 9.6-29.1 K and 1.86-17.34 K, respectively, for clumps within S175A, versus 9.2-16.8 K and 2.65-6.42 K for clumps within S175B. Calculated  excitation  temperature is larger for both regions; $T_{ex}$ varies between 19.1-47.7 K for clumps within S175A and 18.5-29.8 K for S175B.  This temperature range is higher than what we expected for S175B as this region is too distant to be warmed up by the H~II region. The hottest clumps within S175A (C5, C6, C8 and C9) lie at the edge of the H~II region and appear to have been warmed by the star exciting the  H~II region. These clumps  also have higher column densities (105.9-290.4 $\times 10^{20}$ cm$^{-2}$) than the rest of the S175A region (13.28-82.4 $\times 10^{20}$ cm$^{-2}$), and these densities are  also larger than the column densities of the clumps within S175B (33.83-100.9  $\times 10^{20}$ cm$^{-2}$). 

\paragraph{Size and Mass:}

The clumps identified within S175B have larger sizes (0.14-0.34 pc) than  the S175A clumps  (0.12-0.22 pc). It was more difficult to identify the boundary of the clumps within S175B, especially for the colder ones (C3, C4, C7, C8 and C11), because of the extended gas emission with the temperature of the clumps' defined edge temperature ($\sim 5$ K).  Any inaccuracy in boundary 
selection, which highly affects the estimated size of the clump, can produce 
significant uncertainty in the mass estimation. 
It is common in molecular clouds studies to define the edge of the clumps based on a detection level above  the background (see \S \ref{larson}). We were unable to use this boundary selection because of the extended $^{12}$CO(2-1)  emissions in S175B,  which cover $\sim 80\%$ of the mapped region. Therefore we determine clump boundaries  based on the half of the peak temperature. This boundary selection accounts for up to 90\% of the total flux (where the boundary is selected to be the background) of the clearly identified clumps within S175A.

We note that some of the clump masses might be overestimated for the smallest clumps in our sample as the lower limit to our clump size is given by our beam size (0.11 pc at 1.09 kpc) and therefore the masses found for these clumps are upper limits.  We note that less than one third of our clumps are affected by this issue.  

 In addition, saturation and self-absorption, especially in $^{12}$CO(2-1) emission lines which results in measuring larger line widths, is another source of inaccuracy in the mass calculation. This effect leads to a significant overestimation of the Virial mass (up to a factor of 60 for  $M_{Vir}^{12}$ compared to $M_{Vir}^{13}$) and   $M_{Vir}^{12}$ is found to be much larger (up to a factor of 600) compared to the calculated $M_{LTE}$. The $M_{Vir}^{12}$ masses are therefore neglected in the rest of this study.

 $^{13}$CO(2-1) is an optically thinner line and consequently less saturated or absorbed. However,  it is also dependent on the $(\Delta V)^2$ and thus is  highly dependent on  internal dynamics. $M^{13}_{Vir}$  is consistent with $M_{int}$ for S175A, indicating that most of these clumps are close to  Virial equilibrium, but it is noticeably larger than $M_{int}$  for S175B clumps. $M_{LTE}$,  which is independent of the cloud dynamics and directly measures the column density, is the smallest calculated mass. In following section we compare these masses to examine the effect of  the gas internal dynamics on mass estimation.

Reviewing the above, we can see that the properties of the clumps in the  molecular cloud, associated with S175 H~II region, are different from what were expected.  We expected more turbulence and internal motion within S175A due to the proximity of the H~II region.   Indeed, we find that S175A contains the clumps with the highest temperature and the largest column densities (C5 and C6); these clumps lie at the edges of the H~II region and are likely to be material that has been swept up by the expansion of the ionization front  and has then condensed.  However,  S175B displays the wider line profiles and appears more likely to be subject to turbulent motion. Our recent studies of S175B-C1 in $^{12}$CO(3-2) (Azimlu, M$^{\textrm c}$Coey \& Fich, in prep.)  presents strong evidence of an outflow within this clump. We discuss further outflow-driven turbulence and its observable affects  in $\S$\ref{larson}.  Broadening this study to include the third region at the northern part of the cloud, S175C,  would give us insight on the impact of the H~II region on the molecular cloud associated with it.

\subsection{Mass distribution in the S175 molecular cloud}

An FCRAO CO(1-0) Survey of the Outer Galaxy \citep{heyer98} provides us with a further opportunity to investigate the large scale structure and mass within the region. 
This data covers an area much larger than the molecular cloud associated with S175 H~II region which guarantees that we have not missed edges of the cloud, but it has    lower spatial (50$''$) and frequency ($\sim$1 km s$^{-1}$) resolution.

From the data of Heyer et. al. (1998) we find that the molecular material associated with the H~II region is reasonably compact and, with a velocity range of -40 km s$^{-1}$ to -60 km s$^{-1}$,  is distinct from the background emission, which has a $V_{LSR}$ of $\sim 0-20$ km s$^{-1}$. Figure \ref{cgps} shows a map of the molecular cloud, integrated from the FCRAO data between -40 km s$^{-1}$ to -60 km s$^{-1}$ and we calculate the integrated mass in this molecular cloud to be  $\sim 550$ M$_{\odot}$.  The  integrated mass condensed in clumps from JCMT observations within S175A and S175B totals 355 M$_\odot$, so 65\%  of mass in the cloud is in dense clump form.     This is  higher than the typical star formation efficiency ($\sim30$\%, Alves et al. 2007) and is likely a consequence partly of the overestimation of the size of some clumps but also of the overestimation of  $M_{int}^{12}$ due to broadened line profiles.  On the other hand only a fraction of the clump mass may end up in a proto-star but we can not resolve stellar-mass cores within our data . Indeed, the total $M_{LTE}$ for the region is 88 $M_{\odot}$, which is 16\% of the mass in the molecular cloud and closer to the typical value.

It is possible that we have not accounted all the material in the region, or maybe part of  the initial mass  is missed.
Fich \& Rodulph (in prep.) calculated a total mass of   0.4 $M_\odot$ for the ionized gas which accounts only for a very small fraction of the total mass.
We do not have any direct estimation of the atomic hydrogen mass 
but the atomic hydrogen, which lies in a thin layer around the ionized front, cannot contribute  significantly (less than 1\%) to the mass of the cloud \citep{krco}.

The velocity range within the S175 cloud is found to be -47 to -53 km s$^{-1}$.  In comparison, Fich et al. (1990)  measured the ionized gas toward the S175 H~II region to be -55 km s$^{-1}$, indicating that the ionized gas is expanding toward us with a velocity of  $\sim$5 km s$^{-1}$. Assuming that the ionized gas has been expanding with the same velocity to its present size (0.63 pc in diameter) it is $\sim 6\times 10^4$ years old. Since the expansion velocity decreases with time, this age is an upper limit  and swept up material can not have been ejected from the cloud within this short time scale. 
We conclude that the material around the S175 H~II region contain more massive clumps than is typical but cannot, with this data, explain why.

As an aside, it is interesting to note that, contrary to the general belief that massive stars form within Giant Molecular Clouds (GMCs) with typical mass of $10^4-10^5 M_{\odot}$,  here we find  a $\approx$ 10 M$_{\odot}$  B1.5V star has been formed within a relatively small cloud with total mass of only 550 $M_{\odot}$.

\subsection{The effect of line width}

\subsubsection{Line width and mass estimates}  

Some mass estimation methods such as the Virial and $^{12}$CO velocity-integrated masses depend on line profile and width, which are highly sensitive to cloud kinematics.  The LTE analysis directly measures column density and is independent of line profile, and therefore of turbulence and internal dynamics. As a result,  comparing masses estimated by these different methods enables an evaluation of turbulence  and kinematics of the clumps.

\begin{figure}
\epsscale{1}
\plottwo{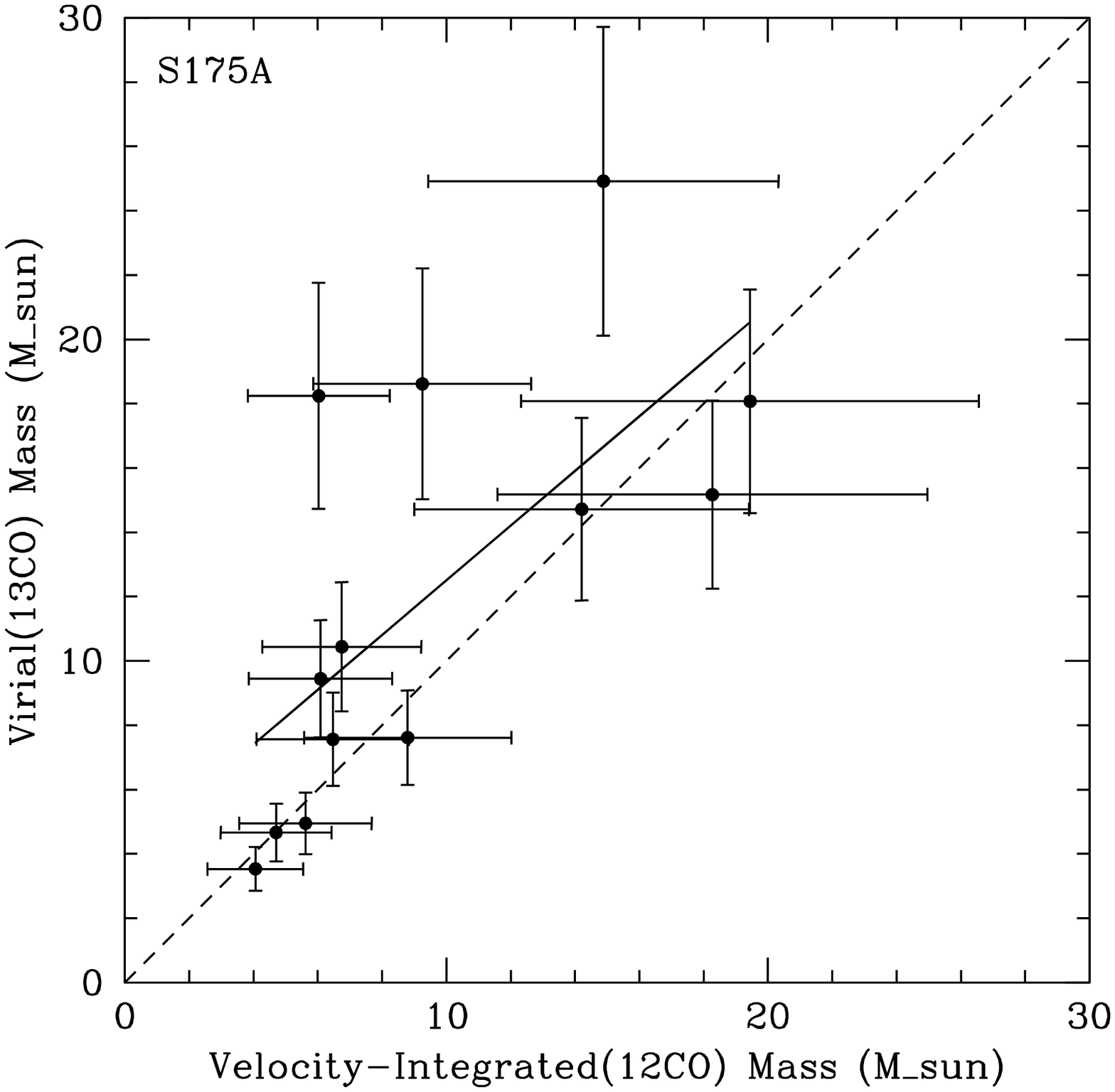}{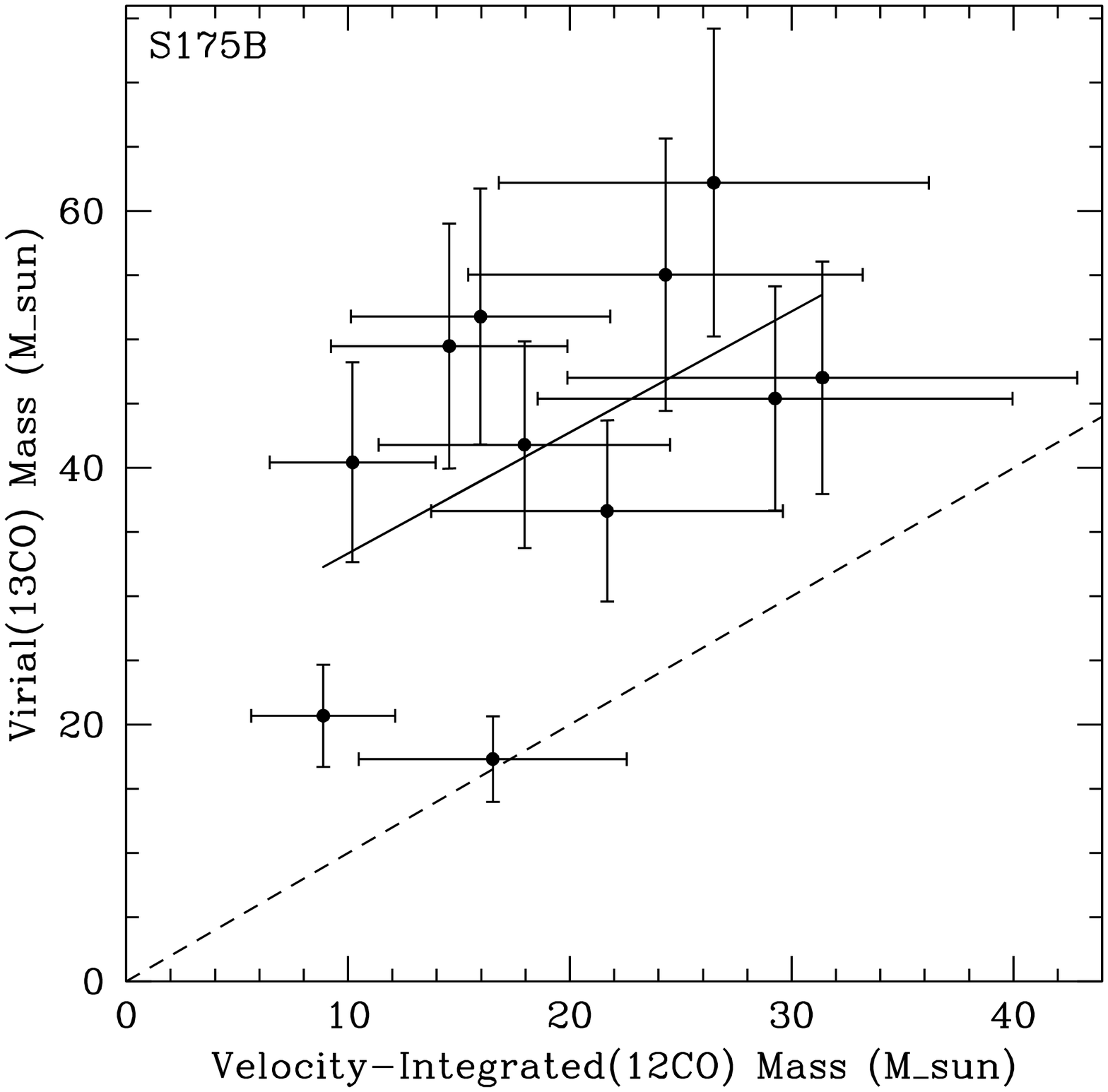}
\caption{ $^{13}$CO Virial mass versus velocity integrated mass, $M^{12}_{int}$ for S175A (left) and S175B (right). 
The solid line is the least square fit and  the long-dashed line indicates the line of equality. Most of the clumps within S175A are in Virial equilibrium. Clumps within  S175B are   dynamically active  and not in a Virial equilibrium.
\label{integvir}}
\end{figure}

In Figure \ref{integvir} we compare $M_{int}^{12}$ and  $M^{13}_{Vir}$. The solid line is the linear least-square fit and the line of equality is marked by long dashes.  In S175A $M^{12}_{int}$ and  $M^{13}_{Vir}$  are almost equal (left panel), indicating that clumps within S175A are close to Virial equilibrium.

The clumps in S175B display larger line widths than those in S175A and have larger velocity-integrated and Virial masses as a result.  The right panel in Figure \ref{integvir} shows there is a weak relation between Virial and velocity-integrated mass in S175B.   
   
An important point to note from this figure  is that S175B is clearly not in Virial equilibrium.

\begin{figure}
\plottwo{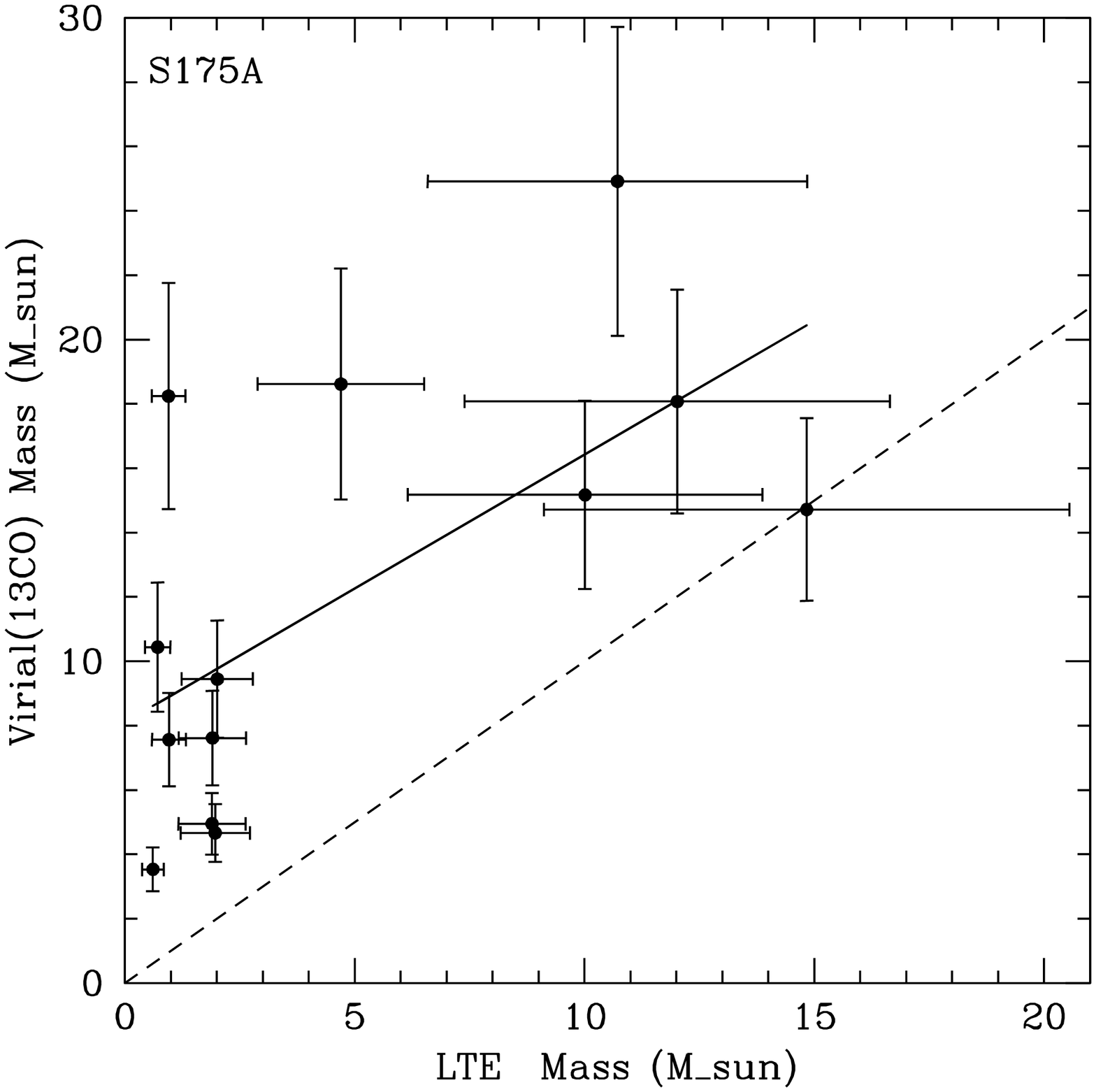}{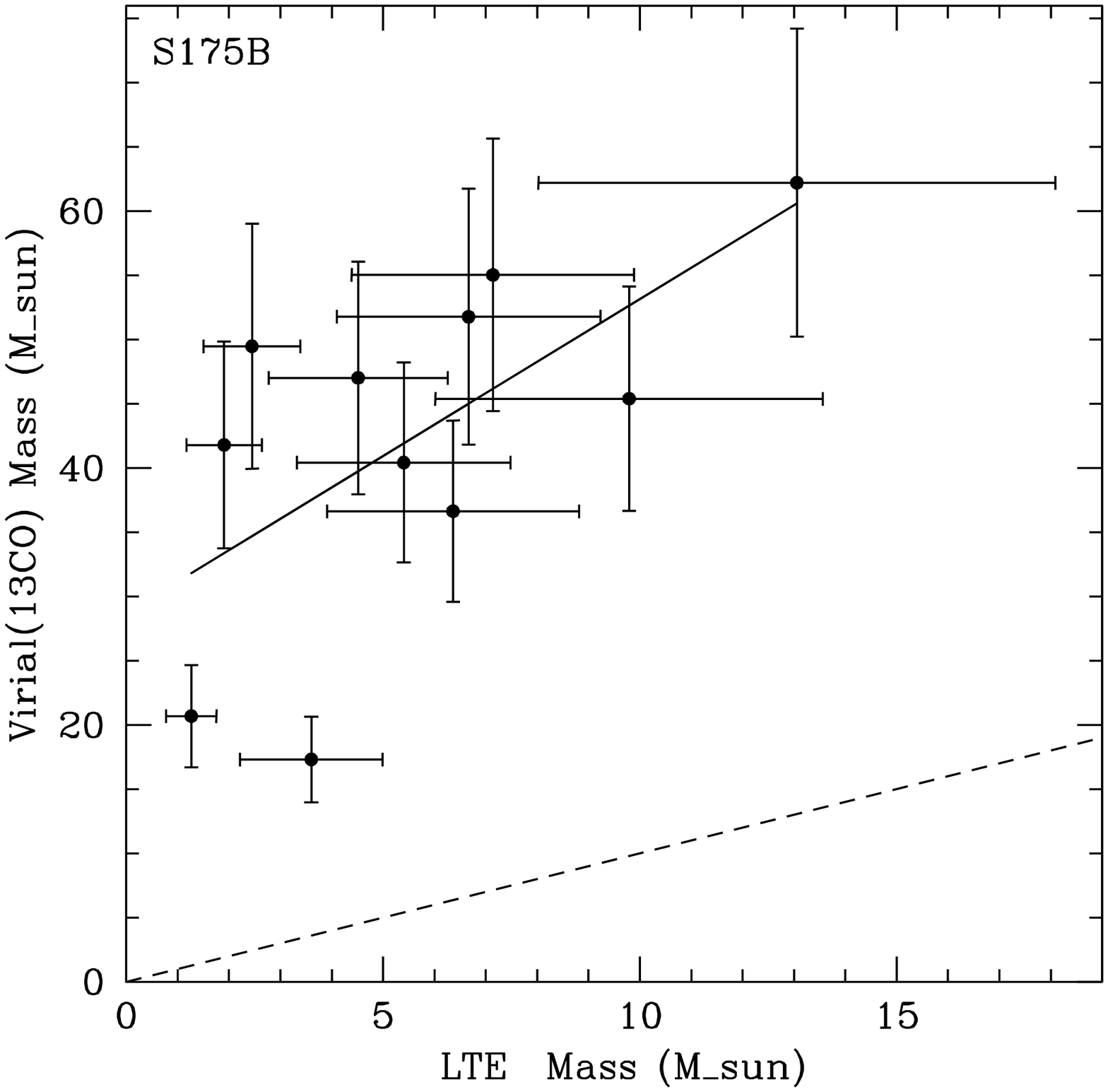}
\caption{  $^{13}$CO Virial mass versus LTE mass for S175A (left) and S175B (right). The dashed line indicates the line of equality.
  Most of the clumps  are over-Virialized compared to LTE mass for both regions but  within S175B $M_{Vir}^{13}$ is noticeably larger than $M_{LTE}$ up to a factor of 20. \label{virlte}}
\end{figure}

\begin{figure}
\plottwo{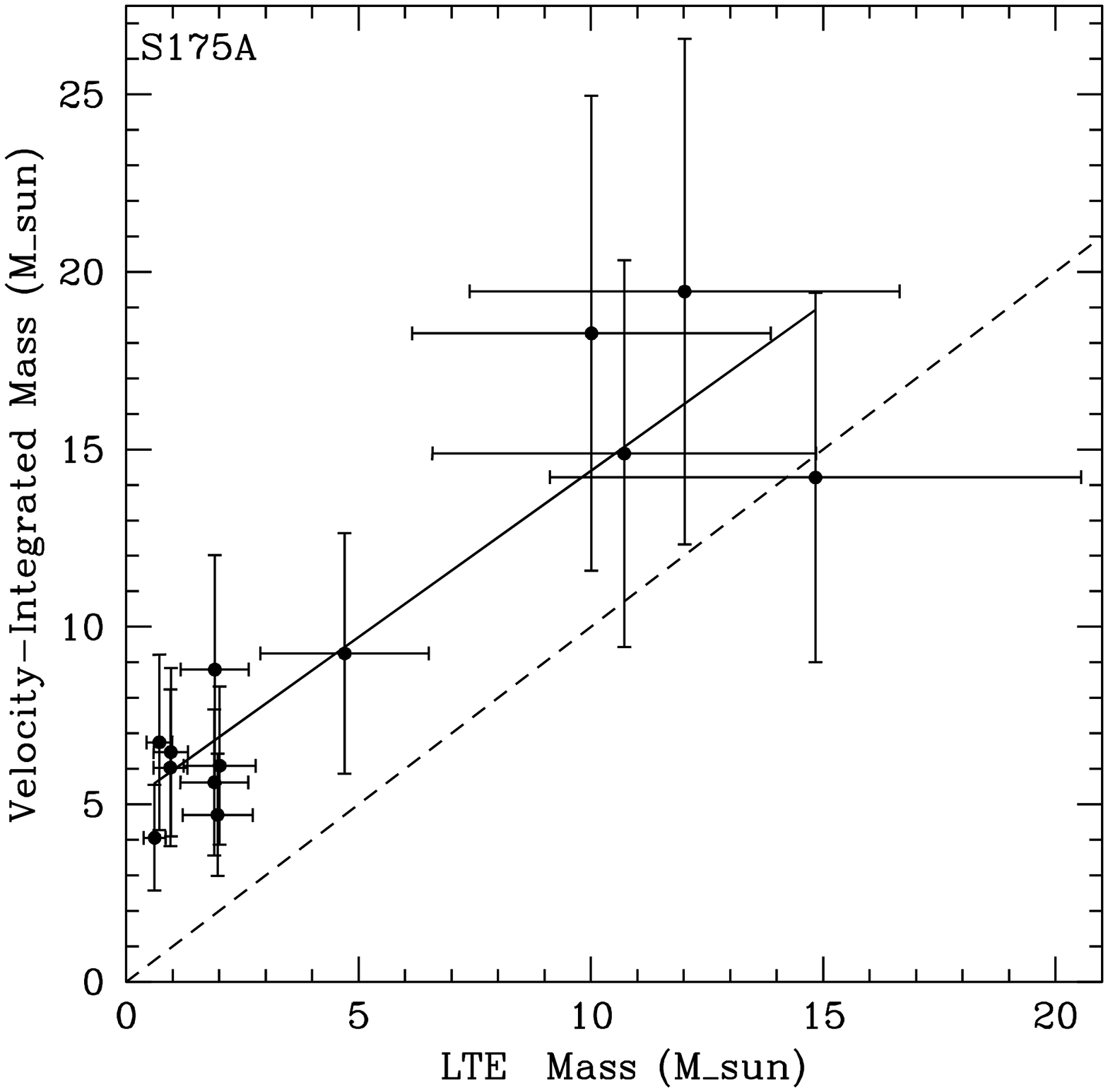}{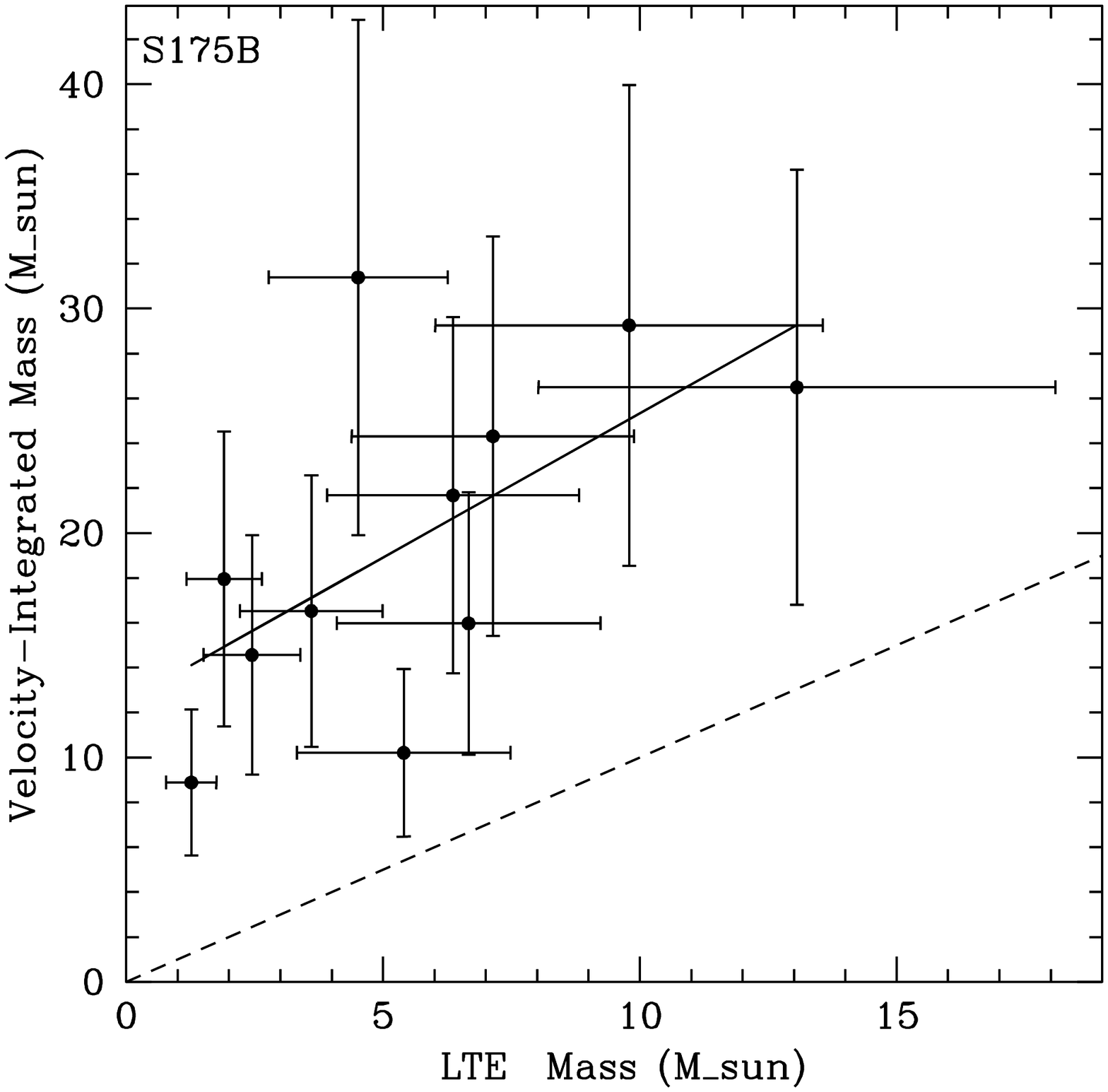}
\caption{Velocity integrated mass versus LTE mass. The dashed line indicates the line of equality. $M^{12}_{int}$  mass is slightly larger than the  $M_{LTE}$ for S175A but  noticeably larger for S175B. \label{integlte}}
\end{figure}

\begin{figure}
\includegraphics[scale=0.7]{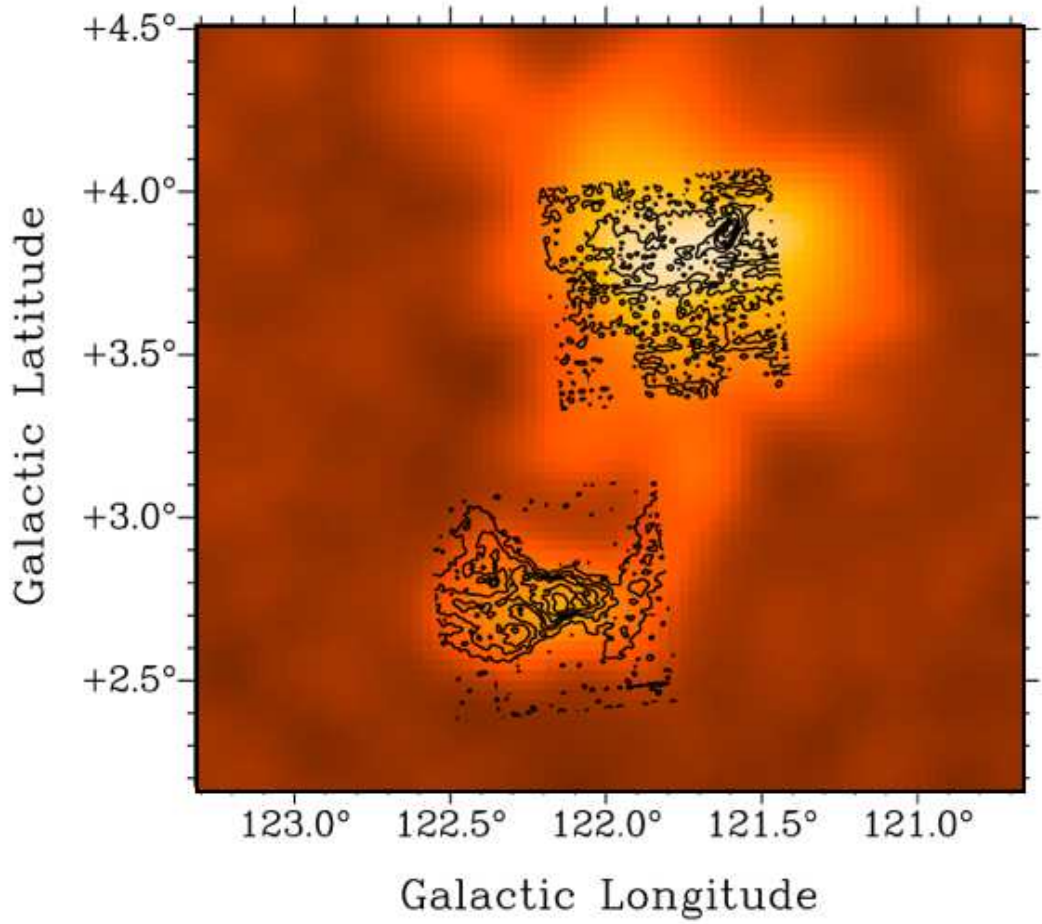}
\caption{FCRAO $^{12}$CO(1-0) map integrated between -40 and -60 km s$^{-1}$. Black contours show the mapped regions in $^{12}$CO(2-1) in our study.  \label{cgps}}
\end{figure}

We can investigate further this issue by plotting the LTE masses against $M_{Vir}^{13}$ (Figure \ref{virlte}) and against $M_{int}^{12}$ (Figure \ref{integlte}).  These Figures show that the LTE masses found are  slightly smaller than either the Virial or integrated-velocity masses for S175A but noticeably smaller for S175B.  In fact, the Virial masses in S175B are up to a factor of 20 larger than $M _{LTE}$.    It is possible that the LTE analysis may underestimate the mass because of self-absorption and saturation of the line profiles, this is especially true in S175B where most of the strongest emission lines are self-absorbed or distorted by internal dynamics, even for $^{13}$CO(2-1).  However,   we note that the ratio of $M_{int}^{12}$ to $M_{LTE}$ for S175B-C1, -C2 and -C3, which show evidence of infall and outflow, is the greatest among all the clumps in the S175 region.  We may conclude that  $M_{int}^{12}$ is overestimated for all dynamically active clumps.

\subsubsection{Line width-clump size relation}
\label{larson}
Various studies of molecular cloud clump/cores have examined the relation between line width and clump/core size (e.g., Larson 1981, Solomon et al. 1987, Lada et al. 1991, Simon et al., 2001) and a power law, often known as the Larson relation, is  reported:

\begin{equation}
\Delta V \propto r^{\alpha},  \hspace{1cm}   0.15 <\alpha< 0.7.
\end{equation}

\noindent This relation has been observed over a large range of clumps/cores, from smaller than 0.1 pc up to larger than 100 pc, which suggests a natural origin.  However, the physical process that results in the relation is still undetermined. 

\begin{figure}
\plottwo{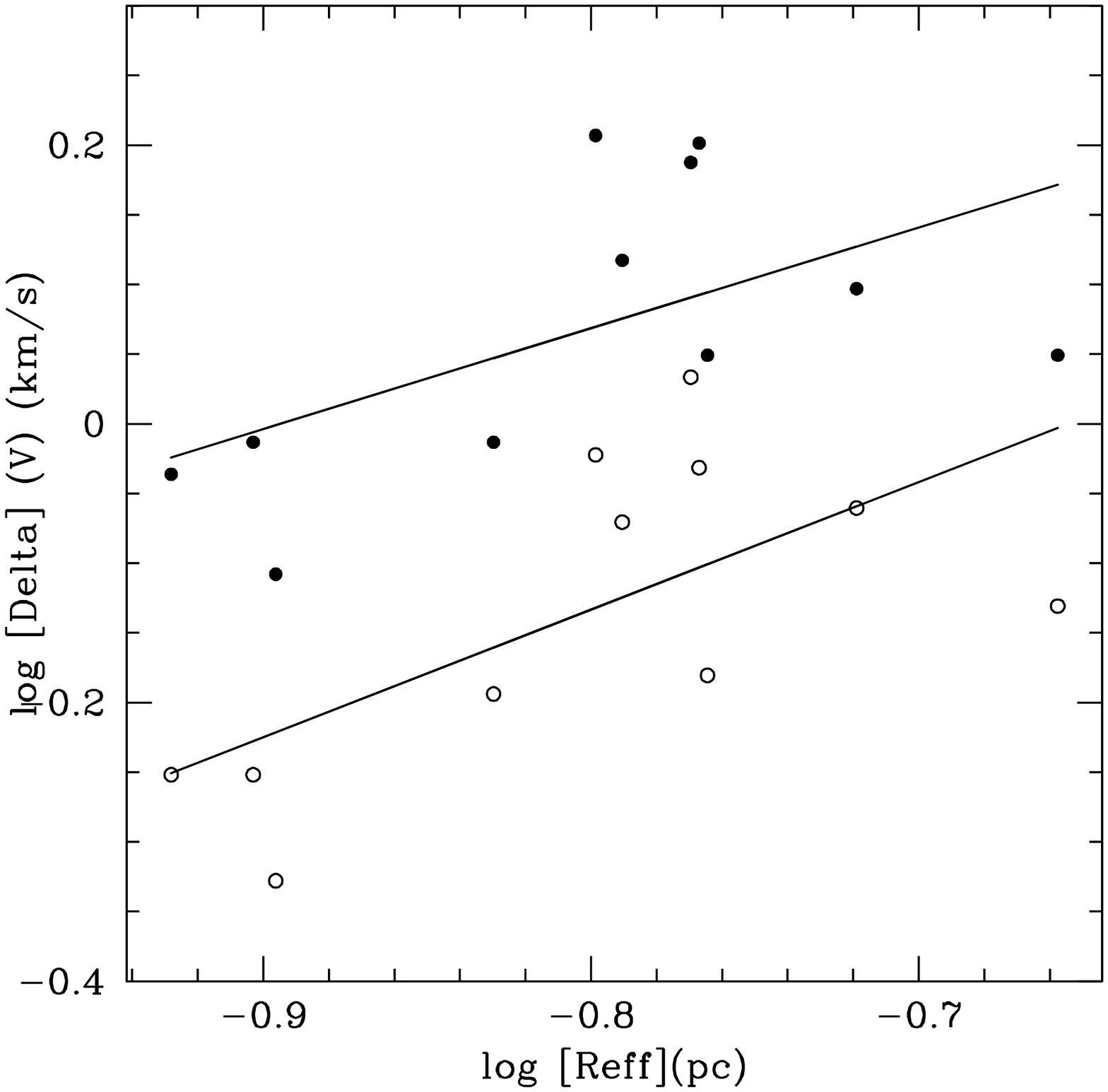}{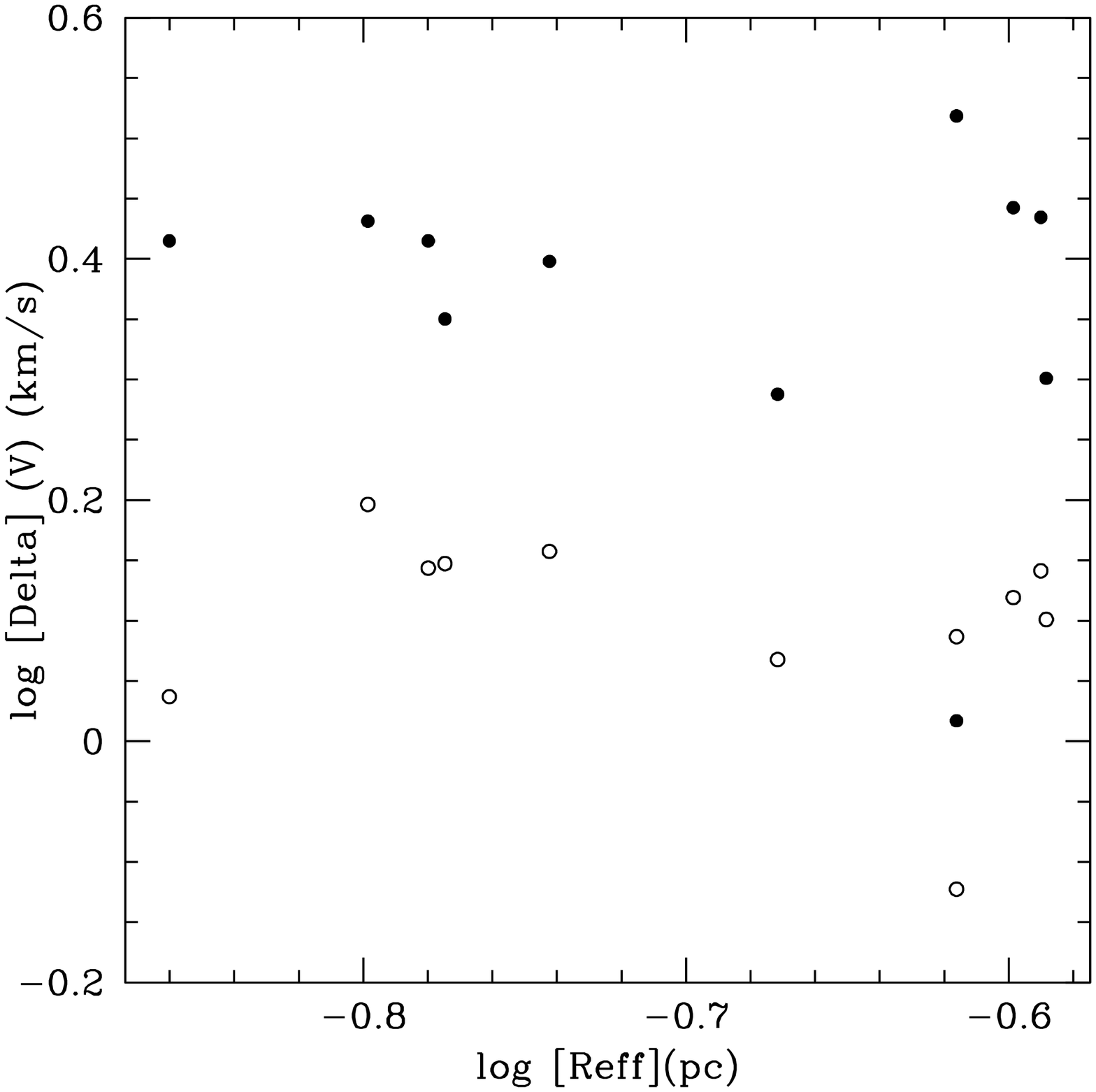}
\caption{ Line width  is weakly correlated with size for S175A clumps (left panel) but no correlation is seen for S175B clumps (right panel).  \label{S175A-4}}
\end{figure}

In Figure \ref{S175A-4} we plot the log of $^{12}$CO(2-1) and  $^{13}$CO(2-1) velocity dispersion versus the effective clump radius in order to investigate the Larson relation in S175.  A least square analysis shows a weak relation in S175A with $\alpha=0.9$ for $^{12}$CO and $\alpha=0.7$ for $^{13}$CO but no correlation is seen for S175B.  
This is not unprecedented: Lada et.al. (1991)  suggested that the existence of the relation is highly affected by the way clumps are defined and selected.  They identified clumps from a survey in the Orion B molecular cloud as having  3$\sigma$ and  5$\sigma$ boundary  above the noise, and found a weak relation for the 5$\sigma$ clumps, but no apparent relation for 3$\sigma$ clumps.  More recent studies in the Galactic plane have also indicated a weak or highly scattered relation \citep{simjack2001}.   Although we base our boundary selection on the peak temperature within a clump rather than on a level above the background, we see a poor correlation  between size and line width  for S175A clumps, however,  no correlation  is found  for S175B.

The possible impact of the broadened line profiles in S175B is intriguing. In a study of cloud cores associated with water masers, Plume et al. (1997)  noted  that the line width-size relation breaks down in massive cores which systematically have higher line widths.  Line widths larger than that expected from thermal motions are thought to be due to local turbulance (Zuckerman \& Evans, 1974). Matzner (2007) modeled outflow-driven turbulence from proto-stars and found that at scales much smaller than the turbulent driving scale the Larson relation is obtained.  At larger scales, however, the line width-size relation flattens in a similar way to that seen in the right panel of FIgure \ref{S175A-4}.   Our recent observations in $^{12}$CO(3-2) of S175B-C1, shows strong signatures of an outflow in this clump (Azimlu, M$^{\textrm c}$Coey \& Fich, in prep.).  The feedback from this outflow is a possible source of turbulence in S175B, at least in the North-Western part of the cloud.  Presence  this outflow indicates that S175B is  an active star forming region and older  outflows of recently formed stars   could be responsible for the observed  structure and kinematics  of this cloud.

\section{Summary and Conclusion}

The CO observations presented here show the presence of a small molecular cloud
 adjacent to the S175 HII region.  We mapped two $7'\times7'$ areas in $^{12}$CO(2-1) within the cloud.  We selected these two regions, S175A and S175B, to investigate the impact of the H~II region on the physical properties of the molecular cloud. Remarkably, these two regions have a very different morphology; S175A is close to the HII region and has a clearly defined, clumpy structure that lies over  the HII region at South-East, while  the more distant S175B  is more uniform but with three clumps that show signatures of outflow.  We studied the clumpy structure of the cloud and identified 13 clumps within S175A and 12 within S175B. We then used pointed observations in $^{13}$CO(2-1) at the brightest peak of   $^{12}$CO(2-1) within each clump in combination with the $^{12}$CO(2-1) spectrum parameters to measure and calculate the physical parameters of the clouds. Our findings are summarized below:

 \begin{enumerate}
 
 \item S175A contains warmer clumps than S175B.
 The most likely explanation for this is that S175A has been heated by the H~II region. However S175B is also hotter than might be expected, probably because of the on-going star formation. 
  
\item  The line widths  are much larger for clumps within S175B than S175A. Comparison of   $\Delta V_{13}$ with thermal line widths

 shows that S175A has not been much disturbed by the H~II region,  while S175B clearly displays non-thermal motions that are most likely the result of an outflow.  In addition to displaying broad line profiles in  $^{12}$CO, the  $^{13}$CO spectra from S175B are split into multiple peaks.
 
 \item 
Clumps within S175B are generally larger.  We suggest that clumps within S175A may be partially a product of the ``collect and collapse" effect of ionization fronts from the H~II region, while a source of turbulence, possibly the outflow, supports the gas within S175B against fragmentation and gravitational collapse. 
 
 \item We determined the  Virial,  LTE and velocity integrated masses  using both $^{12}$CO and $^{13}$CO  lines. $M_{Vir}^{12}$  is over-estimated due to broadened  line profiles and is not included in our analysis.
We find that $M_{Vir}^{13}$ $>$ $M_{int}$ $>$  $M_{LTE}$ in all clumps, which highlights the influence of line width on the mass estimation.  A clear relation between $M_{Vir}^{13}$ and $M_{int}^{12}$ in S175A  indicates that most of the clumps are in Virial equilibrium within this region. No such relation is found for S175B.
 
 \item We investigated the size-line width relationship for clumps and found a weak relation for clumps within S175A. No  relation was found  for S175B clumps, which  may be a result of  a local source of turbulence, due to the outflow, or may be a consequence of the confusion of the clump edges with underly extended gas in S175B. 
  
 \item Observations of massive star forming regions indicate that massive stars originate within Giant Molecular Clouds.  However, the molecular cloud associated with the S175 H~II region is a small, distinct cloud with total mass of 550 M$_\odot$ and it hosts a $\sim 10$M$_\odot$ B1.5V star.  Upto 65 percent of the cloud mass is located within the condensed clumps. 
  
   \end{enumerate}

 We compare the structure and gas properties of these two regions to investigate how the molecular gas has been affected by the H~II region.
 S175A  has been heated by the H~II region and partially compressed by the ionized gas fronts, but contrary to our expectation it  is a quiescent region while S175B is very turbulent and dynamically active. Consequently, we have been able to investigate the influence of turbulence on commonly used mass estimates and line width-size relation. We detected an outflow within S175B which is likely to be the source of turbulence that supports the cloud against gravitational collapse and  causes the observed uniform structure within S175B.

\begin{center}
{\bf Acknowledgment}\\
\end{center}
We thank the anonymous referee for his/her careful reading and constructive remarks. 
\\
The JCMT observatory staff, especially Ming Zhu, Gerald Schieven and Jan Wouterloot are thanked for their support during the observations and their helpful assistance to reduce data. 
The James Clerk Maxwell Telescope is operated by The Joint Astronomy Centre on behalf of the Science and Technology Facilities Council of the United Kingdom, the Netherlands Organisation for Scientific Research, and the National Research Council of Canada.
This research was supported with funding to M.F. from the Natural Sciences and Engineering Council of Canada. 

\clearpage

\clearpage

\clearpage

\clearpage
\vspace{-1 cm}

\clearpage

\clearpage

\clearpage

\clearpage

\clearpage

\clearpage

\begin{deluxetable}{cccccccc}
\tabletypesize{\scriptsize}
\tablecaption{Physical parameters measured for  S175A clumps
\label{tbl-1}}
\tablewidth{0pt}
\tablehead{
\colhead{ID} & \colhead{Position} & \colhead{RA} & \colhead{DEC}&
\colhead{$R_{e}$} & \colhead{$^{12}T_a^{*}$} & \colhead{$^{13}T_a^{*}$}&

\\
 & \colhead{(X,Y)}& \colhead{}&\colhead{}&\colhead{(pc)}&\colhead{(K)}&
\colhead{(K)}&

}
\startdata
C1 & (07,31) &00:27:29.63 & +64:43:41.30 &0.13$\pm$0.024& 19.59 &
 8.41\\
C2 & (08,38) &00:27:27.47 & +64:44:45.21 &0.13$\pm$0.024& 15.86 &
 4.00\\
C3 & (14,47) &00:27:22.02 & +64:45:34.30 &0.17$\pm$0.033& 12.27 &
 5.14\\
C4 & (16,41) &00:27:19.81 & +64:44:52.34 &0.12$\pm$0.023& 18.77 &
 9.45\\
C5 & (30,32) &00:27:04.62 & +64:43:49.51 &0.16$\pm$0.031& 29.09 &
 17.34\\
C6 & (32,29) &00:27:02.30 & +64:43:28.53 &0.16$\pm$0.031& 28.42 &
 14.40\\
C7 & (08,31) &00:27:28.53 & +64:43:42.19 &0.15$\pm$0.028& 20.60 &
 5.75\\
C8 & (20,23) &00:27:15.40 & +64:42:46.60 &0.22$\pm$0.042& 23.22 &
 9.86\\
C9 & (38,29) &00:26:55.74 & +64:43:35.58 &0.17$\pm$0.033& 21.20 &
 11.59\\
C10 &(42,30) &00:26:51.37 & +64:43:35.58 &0.17$\pm$0.033& 14.95 &
 7.41\\
C11 &(47,21) &00:26:41.53 & +64:42:32.59 &0.19$\pm$0.037& 10.80 &
 1.86\\
C12 &(51,30) &00:26:41.53 & +64:43.35.58 &0.16$\pm$0.032& 10.89 &
 2.26\\
C13 &(55,39) &0:26:37.153 & +64:44:38.57 &0.16$\pm$0.030& 9.58 &
 3.61\\
\enddata
\end{deluxetable}

\begin{deluxetable}{cccccccc}
\tabletypesize{\scriptsize}
\tablecaption{Physical parameters measured for  S175B clumps
\label{S175Bt1}}
\tablewidth{0pt}
\tablehead{
\colhead{ID} & \colhead{Position} & \colhead{RA} & \colhead{DEC}&
\colhead{$R_{e}$} & \colhead{$^{12}T_a^{*}$} & \colhead{$^{13}T_a^{*}$}&

\\
 & \colhead{(X,Y)}& \colhead{}&\colhead{}&\colhead{(pc)}&\colhead{(K)}&
\colhead{(K)}&

}
\startdata
C1 & (48,45)&0:26:05.41&+64:54:20.87& 0.18$\pm$ 0.04&12.10&4.90\\
C2 & (35,38)&0:26:19.70&+64:53:32.71& 0.14$\pm$ 0.03&13.81&2.65\\
C3 & (43,34)&0:26:10.90&+64:53:03.86& 0.11$\pm$ 0.02& 9.16&3.54 \\
C4 & (57,19)&0:25:55.52&+64:51:18.84& 0.26$\pm$ 0.05&9.87&4.28 \\
C5 & (37,19)&0:26:17.49&+64:51:18.84& 0.24$\pm$ 0.05&16.82&3.98\\
C6 & (32,32)&0:26:23.00&+64:52:49.81& 0.14$\pm$ 0.03&10.42&3.46 \\
C7 & (22,32)&0:26:33.99&+64:52:49.71& 0.25$\pm$ 0.05& 9.33&4.51 \\
C8 & (24,37)&0:26:31.80&+6:53:24.73& 0.24$\pm$ 0.05&9.53&6.16\\
C9 & (23,46)&0:26:32.92&+64:54:27.72& 0.26$\pm$  0.05&10.71&6.42\\
C10 &(27,43)&0:26:28.51&+64:54:06.77& 0.21$\pm$ 0.04& 12.87& 5.76\\
C11 &(16,50)&0:26:40.63&+64:54:55.63& 0.17$\pm$ 0.03& 10.09&6.34\\
C12 &(7,19) &0:26:50.44&+64:51:18.64& 0.32$\pm$ 0.06&10.38&...\\
\enddata
\end{deluxetable}

\begin{deluxetable}{cccccccccccc}
\tabletypesize{\scriptsize}

\tablecaption{Physical parameters calculated for S175A clumps
\label{tbl-2}}
\tablewidth{0pt}
\tablehead{
\colhead{Clump}&\colhead{$T_{ex}$}& \colhead{$V_{12}$}
 &\colhead{$V_{13}$} &
\colhead{$ \Delta V_{12}$} & \colhead{$ \Delta V_{13}$} &
\colhead{$N(H)$} & \colhead{$n_{int}(H_2)$} &\colhead{$n_{LTE}$}&
\colhead{$\tau_{12}$} & \colhead{$\tau_{13}$ }

\\
\colhead{No.}&\colhead{(K)}&\colhead{$(km s^{-1})$}&\colhead{$(km s^{-1})$}&
\colhead{$(km s^{-1})$} & \colhead{$(km s^{-1})$}&\colhead{$(\times
 10^{20}cm^{-2})$}&
\colhead{$(\times 10^{3}cm^{-3})$}&\colhead{$(\times 10^{3}cm^{-3})$}&
\colhead{}&

}
\startdata
C1&33.83&-49.39&-49.14&0.97&0.56&61.83&6.2&2.3&34.44&0.56\\
C2&28.36&-48.98&-49.03&0.78&0.47&19.38&4.3&0.7&17.85&0.29&\\
C3&23.08&-49.39&-49.03&1.12&0.66&34.68&2.6&1.0&33.22&0.54\\
C4&32.63&-49.39&-49.14&0.92&0.56&72.12&6.2&2.8&42.99&0.69\\
C5&47.68&-49.39&-49.35&1.31&0.85&290.4&7.3&7.9&55.79&0.90\\
C6&46.70&-49.79&-49.67&1.61&0.95&243.4&10.5&6.7&43.52&0.70\\
C7&35.31&-49.39&-49.56&0.97&0.64&44.75&6.0&1.4&20.13&0.32\\
C8&39.13&-50.00&-49.99&1.12&0.74&105.9&3.7&2.2&33.99&0.55\\
C9&36.18&-50.20&-49.67&1.54&1.08&190.8&6.6&5.1&48.59&0.78\\
C10&27.03&-50.81&-50.52&1.59&0.93&82.41&4.0&2.3&41.94&0.68\\
C11&20.90&-50.81&-50.62&1.25&0.87&13.28&1.9&0.3&11.54&0.19\\
C12&21.04&-50.01&-50.84&1.15&0.71&13.50&3.3&0.4&14.21&0.23\\
C13&19.08&-50.01&-50.94&1.32&0.62&20.09&3.7&0.6&28.87&0.47\\
\enddata
\end{deluxetable}

\begin{deluxetable}{ccccccccccc}
\tabletypesize{\scriptsize}

\tablecaption{Physical parameters calculated for S175B clumps
\label{S175Bt2}}
\tablewidth{0pt}
\tablehead{
\colhead{Clump}&\colhead{$T_{ex}$}& \colhead{$V_{12}$}
 &\colhead{$V_{13}$} &
\colhead{$ \Delta V_{12}$} & \colhead{$ \Delta V_{13}$} &
\colhead{$N(H)$} & \colhead{$n(H_2)$} &
\colhead{$\tau_{12}$} & \colhead{$\tau_{13}$ }

\\
\colhead{No.}&\colhead{(K)}&\colhead{$(km s^{-1})$}&\colhead{$(km s^{-1})$}&
\colhead{$(km s^{-1})$} & \colhead{$(km s^{-1})$}&\colhead{$(\times
 10^{20}cm^{-2})$}&
\colhead{$(\times 10^{3}cm^{-3})$}&
\colhead{}&

}
\startdata
C1-a&22.82&-52.03&-51.12&8.68&1.44&70.71&11.6&31.74&0.51\\
C1-b&22.82&...&-49.79&8.68&1.44&70.71&11.6&31.74&0.51\\
C1-c&22.82&-52.03&-48.77&8.68&1.72&70.71&11.6&31.74&0.51\\
C2-a&25.34&-51.62&-51.12&2.24&2.04&34.42&9.6&13.06&0.21\\
C2-b&25.34&-48.98&-49.29&2.77&1.40&34.42&9.6&13.06&0.21\\
C3-a&18.47&-52.23&-51.99&2..70&2.11&49.57&9.3&29.81&0.48\\
C3-b&18.47&-51.22&-50.05&3.11&1.57&49.57&9.3&29.81&0.48\\
C4&19.52&-52.03&-51.22&2.00&1.26&51.17&2.0&34.70&0.56\\
C5&29.77&-49.80&-49.75&1.04&0.75&31.58&2.5&16.57&0.27\\
C6&20.34&-49.39&-49.56&2.60&1.35&33.83&7.3&24.68&0.40&\\
C7-a&18.71&-50.61&-49.57&2.77&1.32&57.44&3.3&40.29&0.65\\
C7-b&18.71&-47.76&-47.73&2.77&1.95&57.44&3.3&40.29&0.65\\
C8&19.02&-50.81&-49.49&3.30&1.22&85.83&4.5&63.01&1.02\\
C9&20.77&-50.61&-49.49&2.72&1.39&100.9&3.4&55.71&0.90\\
C10&23.95&-48.78&-49.22&1.94&1.51&71.87&4.9&36.30&0.59\\
C11&19.84&-48.78&-49.30&2.60&0.59&100.7&4.9&60.15&0.97\\
C12&20.27&-48.78&...&1.13&...&...&0.09&...&...\\
\enddata
\end{deluxetable}

\begin{deluxetable}{cccccccc}
\tabletypesize{\scriptsize}
\tablecaption{Masses calculated for S175A clumps
 \label{tbl-3}}
\tablewidth{0pt}
\tablehead{
\colhead{Clump}& \colhead{$M_{vir}^{13}$} &
\colhead{$M_{int}^{12}$} &
\colhead{$M_{LTE}^{13}$} & \colhead{$M_{Jeans}$} &

}
\startdata
C1&4.9$\pm$1&5.6$\pm$2.1&1.9$\pm$0.7&7.9&\\
C2&3.5$\pm$ 0.7&4.1$\pm$1.5&0.6$\pm$0.2&7.3&\\
C3&9.4$\pm$1.8&6.1$\pm$2.2&2$\pm$0.8&6.9&\\
C4&4.7$\pm$0.9&4.7$\pm$1.7&2$\pm$0.8&7.5&\\
C5&14.7$\pm$2.8&14.2$\pm$5.2&14.8$\pm$5.7&12.2&\\
C6&18.1$\pm$3.5&19.4$\pm$7.1&12$\pm$4.6&9.8&\\
C7&7.6$\pm$1.5&8.8$\pm$3.2&1.9$\pm$0.7&8.6&\\
C8&15.2$\pm$2.9&18.3$\pm$6.7&10$\pm$3.9&12.7&\\
C9&24.9$\pm$4.8&14.9$\pm$5.4&10.7$\pm$4.1&8.4&\\
C10&18.6$\pm$3.6&9.3$\pm$3.4&4.7$\pm$1.8&7.0&\\
C11&18.2$\pm$3.5&6.0$\pm$2.2&0.9$\pm$0.4&7.0&\\
C12&10.4$\pm$2.0&6.7$\pm$2.5&0.7$\pm$0.3&5.3&\\
C13&7.6$\pm$1.5&6.5$\pm$2.4&1$\pm$0.4&4.3&\\
\enddata
\end{deluxetable}

\begin{deluxetable}{ccccccc}
\tabletypesize{\scriptsize}
\tablecaption{Masses calculated for S175B clumps \label{S175Bt3}}
\tablewidth{0pt}
\tablehead{
\colhead{Clump}& \colhead{$M_{vir}^{13}$} &
\colhead{$M_{int}^{12}$} &
\colhead{$M_{LTE}^{13}$} & \colhead{$M_{Jeans}$} &

}
\startdata
C1&47$\pm$9.1 &31$\pm$11.5&4.5$\pm$1.7&3.2&\\
C2&42$\pm$8.1&18$\pm$6.6&1.9$\pm$0.7&4.5&\\
C3&50$\pm$9.5&15$\pm$5.3&2.4$\pm$0.9&2.8&\\
C4&52$\pm$10&16$\pm$5.8&6.7$\pm$2.6&6.1&\\
C5&17$\pm$3.3&17$\pm$6.0&3.6$\pm$1.4&10.2&\\
C6&21$\pm$4.0&9$\pm$3.3&1.3$\pm$0.5&3.4&\\
C7&55$\pm$10.6&24$\pm$8.9&7.1$\pm$2.8&4.5&\\
C8&45$\pm$8.7 &29$\pm$10.7&9.8$\pm$3.8&3.9&\\
C9&62$\pm$12&27$\pm$9.7&13.1$\pm$5.0&5.1&\\
C10&37$\pm$7.1&22$\pm$7.9&6.4$\pm$2.5&5.3&\\
C11&40$\pm$7.8&10$\pm$3.7&5.4$\pm$2.1&4&\\
C12&...&12$\pm$4.5&...&9.8&\\

\enddata
\end{deluxetable}

\end{document}